\documentclass{article}
\pdfoutput=1
\usepackage{arxiv}
\usepackage{float}
\floatstyle{plaintop}
\restylefloat{table}
\usepackage{amssymb}
\usepackage{amsmath,amsfonts}
\usepackage{tcolorbox}
\usepackage{tikz}
\usetikzlibrary{shapes.geometric, arrows}
\usepackage{ mathrsfs }
\usepackage{listings}
\usepackage{color}
\usepackage[caption=false]{subfig} 
\usepackage{blindtext}
\usepackage{setspace}
\usepackage{gensymb}
\usepackage{pgfplots}
\pgfplotsset{compat=1.8}
\usepackage{mathtools}
\usepackage{stmaryrd,dsfont}
\usepackage{bbm}
\usepackage{url}
\usepackage{wrapfig,bm}
\usepackage[export]{adjustbox}
\usepackage{pgfplots}
\usepackage{tikz-3dplot}
\usepackage{booktabs}
\usepackage{lineno}
\usepackage{url}
\usepackage{listings}
\usepackage{xcolor} % for setting colors
\usepackage[ruled,vlined]{algorithm2e}
\newcommand{\eref}[1]{Equation~(\ref{#1})}
\newcommand{\erefm}[1]{Eqn.~(\ref{#1})}

\newcommand{\fref}[1]{Figure~\ref{#1}}

\newcommand{\frefs}[1]{Figures~\ref{#1}}

\newcommand{\vm}[1]{\bm{\mathrm{#1}}} % this makes straight bold letters

\definecolor{forestgreen}{RGB}{255,100,100}

\title{Continuum modelling of stress diffusion interactions in an elastoplastic medium in the presence of geometric discontinuity}
\author{Rupesh Kumar Mahendran \\
Mechanics of Materials Laboratory, \\Department of Mechanical Engineering,\\ Indian Institute of Technology Madras,\\ Chennai-600036, India. \\ \texttt{dmrupeshkumar@gmail.com} \\
\And Hirshikesh Hirshikesh 
\\
Mechanics of Materials Laboratory, \\Department of Mechanical Engineering,\\ Indian Institute of Technology Madras,\\ Chennai-600036, India. \\ \texttt{me15d404@smail.iitm.ac.in}\\ \And Sundararajan Natarajan \\
integrated Modeling and Simulation Lab, \\Department of Mechanical Engineering,\\ Indian Institute of Technology Madras,\\ Chennai-600036, India. \\ \texttt{snatarajan@iitm.ac.in}\\ \And Ratna Kumar Annabattula\\
Mechanics of Materials Laboratory, \\Department of Mechanical Engineering,\\ Indian Institute of Technology Madras,\\ Chennai-600036, India. \\ \texttt{ratna@iitm.ac.in}}
\begin{document}
\maketitle
% \begin{frontmatter}

% %% Title, authors and addresses
% %\title{Continuum modelling of stress diffusion interactions in an elastoplastic medium in the presence of geometric discontinuity}
% %\title{Role of the plastic yielding and the singularity on stress-diffusion interactions and application in particles with Core-Shell structures for for Lithium-Ion battery electrodes}
% \author{Rupesh Kumar Mahendran}
% \author{Hirshikesh Hirshikesh}
% \author{Sundararajan Natarajan\corref{cor1}}
% \ead{snatarajan@iitm.ac.in;snatarajan@cardiffalumni.org.uk}
% \author{Ratna Kumar Annabattula\corref{cor1}}
% \ead{ratna@iitm.ac.in}
% \cortext[cor1]{Corresponding author}
% \address{Mechanics of Materials Laboratory, \\Department of Mechanical Engineering, Indian Institute of Technology Madras, Chennai-600036, India }
\begin{abstract}
 Chemo-mechanical coupled systems have been a subject of interest for many decades now. Previous attempts to solve such models have mainly focused on elastic materials without taking into account the plastic deformation beyond yield, thus causing inaccuracies in failure calculations. This paper aims to study the effect of stress-diffusion interactions in an elastoplastic material using a coupled chemo-mechanical system. The induced stress is dependent on the local concentration in a one way coupled system, and vice versa in a two way coupled system. The time-dependent transient coupled system is solved using a finite element formulation in an open-source finite element solver FEniCS. This paper attempts to computationally study the interaction of deformation and diffusion and its effect on the localization of plastic strain. We investigate the role of geometric discontinuities in scenarios involving diffusing species, namely, a plate with a notch/hole/void and particle with a void/hole/core. We also study the effect of stress concentrations and plastic yielding on the diffusion-deformation. The developed code can be from \textcolor{blue}{https://github.com/mrupeshkumar/Elastoplastic-stress-diffusion-coupling} 
 \end{abstract}


% \end{frontmatter}

% \section*{Highlight}

% \begin{itemize}
%     \item Implemented and validated the transient elasto-plastic coupled stress-diffusion model in FEniCS finite element solver.
%     \item Studied the influence of diffusion/deformation on plastic yielding and vice-versa through one-way and two-way coupled models.
%     \item Investigated the interaction of the coupled diffusion deformation in the presence of geometric discontinuities.
%     \item On comparing one-way and two-way coupled models we observed the relaxation of tensile stress concentration due to two-way coupled diffusion.
%     \item The influence of plastic yielding reduces the local stress concentration leading to a reduction in the local diffusion of species in case of tensile stress and an increase in diffusive species in the compressive region with the introduction of a geometric discontinuity.
%     \item For a particle model we observed reduction in stress singularity leading to a reduction in plastic yielding in tensile region and an increase in plastic yielding in compressive region in the presence of a void.
% \end{itemize}

\section{Introduction}
\addcontentsline{toc}{chapter}{Introduction}

Diffusive species migration, such as hydrogen in steel \cite{Johnson1957}, lithium in Lithium-Ion batteries (LIBs) \cite{Tsuyoshi2013569}, chlorine in cement \cite{PAGE1981395} to name a few, cause local expansion/contraction of the materials leading to the localized chemical strains in the material. If the appropriate deformation of the solid does not accommodate these localized chemical strains, stresses referred to as diffusion-induced stress (DIS) \cite{Lei2016} are induced. In addition to this, the localized stresses act as a driving force for the species diffusion in the materials, which is commonly referred to as stress-induced diffusion (SID) \cite{Simon1992}. Therefore, there is a strong coupling between the diffusion and mechanics of the material, i.e., diffusion of the species causes the localized stresses, and the localized stresses also influence the diffusion of the species and vice-versa. The response of the stress-diffusion interactions can be investigated in two aspects as: 
\begin{itemize}
    \item one-way coupling: where the gradient of concentration affects the deformation field, but the localized stress state does not influence the evolution of concentration,
    \item two-way coupling: in addition to the influence of the gradient of local hydrostatic stress on concentration and the localized concentration also influences the localized stress state.
\end{itemize}

The presence of diffusion species, either as an external gas or resulting from electrochemical reactions leads to the severe degradation in the material properties and failure of the material~\cite{YU201656, Rhode2016201, Feifei2016, Tavassol2016}. Frailure due to these multi-physical phenomenon (stress-diffusion interactions~\cite{FISCHER2015427}) shares a major portion of material failures, viz., chloride diffusion in civil structures~\cite{PAGE1981395, WU2016196, WANG200958}, hydrogen diffusion in steels~\cite{Johnson1957, Oriani1972848,  Hirth1980861, Adrover2003,  Ahn2007,Olden2008, Sobotka2009, song2012}, aluminum~\cite{YOUNG1998}  and NiTi alloys~\cite{Schmidt19892473,Moitra2011820,Letaief2017075001}, bacteria diffusion in biofilms~\cite{philip2003, Duddu200992, Duddu2008848}, Li diffusion in LIBs~\cite{Tavassol2016, Feifei2016,onewaySZH2012, William2010}.

% The study of mechanical degradation due to stress concentrations has been a subject of interest with many models proposing the use of a thermal expansion analogy for thermo-elasticity \cite{Zhang01102007, stein20143d,Yingjie20140299, Swaminathan01012016} to model DIS. 
A number of frameworks for diffusion induced deformation and fracture have been proposed assuming an elastic medium \cite{DESHPANDE20105081, Haftbaradaran2010, Zhang2007, hu_zhao_suo_2010, Christensen2006, AIFANTIS2007874, Woodford01102010, BHANDAKKAR20101424, Wu201719022} and taking thermal analogy to model stress-diffusion interactions \cite{Zhang01102007, stein20143d,Yingjie20140299}. In the design of the system where these interactions involve, the radius and the aspect ratio of the particle plays a crucial role in the concentration as well as the fracture behavior \cite{Zhang01102007, onewaySZH2012}. Furthermore, the fully coupled stress-diffusion interactions and the higher-order terms in the chemical potential also affects the concentration and the stress profile \cite{Swaminathan01012016}. The geometrical, as well as the material discontinuities present in the system, also play a critical role \cite{Natarajan2016, Yingjie20140299}. However, the assumption of an elastic medium may lead to inaccuracies in failure calculations beyond the elastic regime.

Study of failure of metallic specimens containing stress concentrations such as notches and sharp cracks showed that failure happened with restriction of a plastic zone to very near the fracture surface \cite{PERNG19893393, DELAFOSSE2001693, BOND19882193, Fuqian2005}. The hydrogen diffusion induced localized plasticity models \cite{BARRERA2016219, delafosse2009hydrogen, BIRNBAUM1994191, Abraham1995, STASHCHUK201214687, SOFRONIS1989317} were able to show increased dislocation mobility in the presence of hydrogen diffusion, causing embrittlement. In-situ studies by Sethuraman {et. al., }\cite{SETHURAMAN20105062} have shown that the stress exceeds yield strength during large deformation of silicon electrodes due to lithium insertion accompanied by plastic deformation. The plastic deformation helps maintain good performance and capacity during cyclic operation by reducing stress buildup \cite{BRASSART20131120}. Sethuraman {et. al., } \cite{SETHURAMAN20105062} experimentally found the evolution of flow stress induced by lithiation and found a strong hysteresis which showed plastic deformation of the electrode.  Motivated by these observations continuum theories \cite{BOWER2011804,Zhao2011s226,Zhao2011016110, ZHANG201547} have shown diffusion induced stress due to elastic-plastic deformation of the electrode material. 
%\citet{Zhaonl201501s} used first-principles calculations to study the microscopic mechanism of large plastic deformation due to lithium insertion.  

However, the effect of the plasticity on the one way and fully coupled stress diffusion interaction in the presence of the geometric discontinuities are still unclear. In this work, we numerically investigate the role of plasticity in one-way and two-way coupled stress-diffusion interactions. The investigation is performed for the general case of the stress-diffusion interaction framework. Hence, we have performed studies on hydrogen diffusion in steel as well as lithium diffusion in graphite anode particle.

The manuscript is structured as follows: Section \ref{S:2} presents governing equations for the coupled stress-diffusion interaction in an elastoplastic medium and the numerical implementation details in an open-source finite element package FEniCS. Section \ref{S:3} describes the boundary value problem. Section \ref{S:4} presents the results and discussions pertaining to the chosen boundary value problem. The major conclusions are discussed in the last section.

% A significant amount of effort has been made towards modeling the different aspects of the LIB operation. This includes the electrochemical process, stress-induced effects on diffusion, fracture and failure mechanisms. The role of surface stresses on the DIS were analyzed in Refs.[12–14]. DIS can be modeled using a framework similar to coupled thermo-elasticity by treating the chemical and the thermal expansions analogy.[15–17] In this framework, the gradient of the hydrostatic stress field contributes as a drifting term to the classical diffusion equation and hence governs the SID. This drifting term is derived from the chemical potential[16, 18,19] and depends on the gradient of the hydrostatic stress in the body.\textit{Zhang et al.} studied the dilation induced stress in the cathode particle using a three-dimensional finite element model. From the numerical simulation, it was observed that the internal stress gradients significantly enhance the diffusion. Most of the efforts in studying the chemo-mechanical models have been using a linear elastic medium. This may lead to inaccuracies in failure calculations beyond the elastic regime. Therefore this project aims to better understand the effects of two-way coupled stress-diffusion interactions in an elasto-plastic material.  

\label{S:1}

\section{Numerical formulation}
\label{S:2}

In the present study we make use of a continuum model to describe the coupled diffusion deformation. We make an approximation that the diffusion in the medium and the subsequent expansion/contraction is isotropic. Consider an elasto-plastic homogeneous isotropic body occupying an open domain $\beta \subset \mathbb{R}^2$ and bounded by a surface $\Gamma$ with unit outward normal \textbf{n}. Let \textbf{u} : $ \beta \rightarrow \mathbb{R}^2 $ be the displacement field at any point \textbf{x} of the body when the body is subjected to external tractions T : $ \Gamma_t \rightarrow \mathbb{R}^2 $ and body forces \textbf{b} : $\beta \rightarrow \mathbb{R}^2$ and $c\ :\ \beta \rightarrow \mathbb{R}^2$ be the concentration field at a point \textbf{x} when the body is subjected to external flux $J\ :\Gamma_J \rightarrow \mathbb{R}^2 $. The boundary is assumed to admit the following decompositions:
\begin{itemize}
    \item For the elastic-plastic problem: $\Gamma = \Gamma_u \cup \Gamma_t$ and $\emptyset = \Gamma_u \cap \Gamma_t$, where $\Gamma_u$ is the Dirichlet boundary and $\Gamma_t$ is the Neumann boundary and
    \item For the concentration problem: $\Gamma = \Gamma_c \cup \Gamma_J$ and $\emptyset = \Gamma_u \cap \Gamma_t$, where $\Gamma_c$ is the Dirichlet boundary and $\Gamma_J$ is the Neumann boundary
\end{itemize}

The chemo-mechanical problem now simplifies to two independent variables: displacements and concentration.The boundary value problem for the coupled diffusion-deformation model in the absence of body force and the source term then becomes:  find $\boldsymbol{u}:\beta \subset \mathbb{R}^d$ and $c : \beta \subset \mathbb{R}^d$.

The diffusion of the species can be defined by:
\begin{subequations}
\begin{equation}
    \frac{\partial c}{\partial t} + \mathbf{\nabla  \cdot J} = 0,
\label{eq1}   
\end{equation}

and the mechanics of the elasto-plastic domain can be defined b:
\begin{equation}
\label{eq2}
    \nabla \cdot \mathbf{\sigma} = \mathbf{0},
\end{equation}

subject to the following boundary conditions:

\begin{eqnarray}
     u = \hat{u} \hspace{5mm} \forall x \in \Gamma_u,  \\
     c = \hat{c} \hspace{5mm} \forall x \in \Gamma_c, \\
   \bm{ \sigma} \cdot \mathbf{n} = \hat{t} \hspace{5mm} \forall x \in \Gamma_t,  \\
    \mathbf{J} \cdot \mathbf{n} = \hat{\rm{J}} \hspace{5mm} \forall x \in \Gamma_J,  
\end{eqnarray}
\end{subequations}

\noindent where $\boldsymbol{\sigma}$ is the Cauchy stress tensor and $\mathbf{J}$ is the flux. The thermal analogy from previous studies 
%(\cite{zhang2007numerical}, \cite{swaminathan2016elasticity}) 
is used to find the influence of concentration on the stress field as: 
\begin{equation}
\label{eq8}
    \boldsymbol{\sigma} = \mathbb{C}\bigg(\boldsymbol{\varepsilon} - [c-c_0]\frac{\overline{\Omega}}{3} - \boldsymbol{\varepsilon}^p\bigg),
\end{equation}

\noindent and the flux $J$ is related to the concentration and stresses by: 
\begin{equation}
\label{eq9}
   {\rm{J}} = -D\bigg(\nabla c - \frac{\overline{\Omega}}{RT}c\nabla\sigma_h\bigg),
\end{equation}
where $\mathbb{C}$ is the constitutive matrix for an linear isotropic  material, $\bm{\varepsilon}$ is the total infinitesimal strain, $\bm{\varepsilon}^p$ is the plastic strain, $c_0$ is
the initial or the reference concentration, $\overline{\Omega}$ is the partial molar volume, $D$ is the diffusion coefficient, $R$ is the
universal gas constant, $T$ is the absolute temperature and $\sigma_h = \frac{1}{3}\sigma_{ii}$ is the hydrostatic stress.

\subsection{Continuum mechanics modelling}

\noindent In case of a coupled chemo-mechanical system the total strain increment $\dot{\varepsilon}$ can be decomposed additively. Following a small strain assumption proposed by Zhao {\it et. al.,} \cite{Zhao2011016110} the decomposition of the strain increment can be written as,
\begin{equation}
   \mathrm{d} \bm{{\varepsilon}} = \mathrm{d}\bm{{\varepsilon}}^e + \mathrm{d}\bm{{\varepsilon}}^c + \mathrm{d}\bm{{\varepsilon}}^p
\end{equation}
where $\mathrm{d}\bm{{\varepsilon}}^e, \mathrm{d}{\bm{\varepsilon}}^c, \mathrm{d}{\bm{\varepsilon}}^p$ are the increment in elastic strain, diffusional strain and plastic strain respectively.
\noindent The strain-displacement relationship gives,
\begin{equation}
    {\varepsilon}_{ij} = \frac{1}{2}(u_{i,j}+u_{j,i}).
\end{equation}
In case of diffusion induced stress there is an additional strain due to concentration. In case of general isotropic continuum material this is written as a dilational strain rate,

\begin{eqnarray}
    \boldsymbol{\dot{\varepsilon^c}} = \dot{c}\ \frac{\overline{\Omega}}{3}\boldsymbol{I}
\end{eqnarray}
Since the material is elastically isotropic,

\begin{equation}
     \dot{\bm{\sigma}} = \mathbb{C}^e:\dot{\bm{\varepsilon^e}} = \mathbb{C}^e:(\dot{\bm{\varepsilon}}-\dot{\bm{\varepsilon}}_p-\dot{\bm{\varepsilon}}_c)
\end{equation}

\begin{equation}
    \bm{\dot{\varepsilon}}^e = \frac{1+\nu}{E}\dot{\bm{\sigma}} - \frac{\nu}{E}(\dot{\bm{\sigma}}_{kk})\mathbf{I}.
\end{equation}
\noindent The isotropic $J_2$- flow rule gives the yield condition as,
\begin{equation} 
    f = \sigma_e - \sigma_y(\varepsilon^p) = 0
\end{equation}

\noindent where $\sigma_e = \sqrt{\frac{3}{2}\mathbf{S}:\mathbf{S}} $ is the equivalent stress, and $\sigma_y$ is the yield stress, $\mathbf{S}$ is the deviatoric stress. 
The associated flow rule gives the plastic strain rate as,
\begin{equation}
    \dot{\bm{\varepsilon}^p} = \dot{\lambda}\frac{\partial f}{\partial\bm{\sigma}} = \dot{\lambda}\frac{3\mathbf{S}}{2\sigma_e}
\end{equation}

\noindent where $\dot{\lambda}$ is the loading parameter is found using the \textit{consistency condition} $\dot{f}=0$.

The consistency condition is written as:

% \begin{equation} 
%     f = \frac{3}{2}\mathbf{S}:\mathbf{S} - \sigma_y^2(\varepsilon^p) = 0
% \end{equation}

\begin{equation} 
    \dot{f} = \frac{3}{2}\frac{\mathbf{S}:\mathbf{\dot{\sigma}}}{\sigma_e} - (\frac{\partial\sigma_y}{\partial\varepsilon^p_e})\dot{\varepsilon^p_e} = 0
\end{equation}

\begin{equation} 
\dot{\varepsilon^p_e}  =\frac{3}{2} \frac{\mathbf{S}:\mathbf{\dot{\sigma}}}{(\frac{\partial\sigma_y}{\partial\varepsilon^p_e}) \sigma_e} = \dot{\lambda}
\end{equation}

From the plastic flow rule, we get 

\begin{equation} 
\bm{\dot{\varepsilon^p}} = \frac{9}{4} \frac{\mathbf{S}:\mathbf{\dot{\sigma}}}{(\frac{\partial\sigma_y}{\partial\varepsilon^p_e}) \sigma_e^2}  \bm{S}
\end{equation}

\noindent where $\frac{\partial\sigma_y}{\partial\varepsilon^p_e} = H$ is the hardening constant for linear hardening. For numerical simplicity in the context of this paper we have implemented linear isotropic and kinematic hardening to study the evolution of plastic strain fields. For small range of plastic strains it is a reasonable assumption because the focus of this paper is mainly towards studying the influence of plastic strain in SID in the presence of geometric discontinuities. Appendix B contains the detailed derivation for the Kinematic hardening model used for numerical studies in this paper.\\

% \eref{eq8} can be rewritten as 
% \begin{equation}
%     \boldsymbol{d\sigma} = \mathbb{C}:(\boldsymbol{d\varepsilon} - \boldsymbol{d\varepsilon^c} - \boldsymbol{d\varepsilon^p})
% \end{equation}
\subsection{Weak form and finite element implementation}
% The corresponding weak formulation of \eref{eq1} and \eref{eq2} is to find $\boldsymbol{u} \in \boldsymbol{\mathscr{U}}$ and $c \in \mathscr{C}$ such that $\forall \boldsymbol{v} \in \boldsymbol{\mathscr{V}}$ and $\forall p \in \mathscr{P}$:
% \begin{eqnarray}
% \label{eq10}
%     \int_\Omega\bigg(\mathbf{\nabla \cdot \ \sigma}\bigg) \boldsymbol{v}~ \mathrm{d}V = 0, \nonumber \\
% \label{eq11}
%     \int_\Omega  \bigg(\frac{\partial c}{\partial t} + \mathbf{\nabla \cdot J}\bigg)p ~\mathrm{d}V = 0,
% \end{eqnarray}
% \begin{subequations}
% \begin{equation}
%     \int_\Omega\bigg(\mathbf{\nabla \cdot \ \sigma}\bigg) \boldsymbol{v}~ \mathrm{d}V = 0,  
%     \label{eq10}
% \end{equation}
% \begin{equation}
%     \int_\Omega  \bigg(\frac{\partial c}{\partial t} + \mathbf{\nabla \cdot J}\bigg)p ~\mathrm{d}V = 0,
%         \label{eq11}
% \end{equation}
% \label{eqn:strongForm}
% \end{subequations}

Consider $\boldsymbol{\mathscr{U}}$ and $\mathscr{C}$ are the displacement and the concentration trial spaces and $\mathscr{V} $and$ \mathscr{P}$ are the displacement and the concentration test spaces:
$$\mathscr{U} := \{ \textbf{u(x)} \in [c^o(\Omega) ]^d: \textbf{u} \in [\mathscr{W}(\Omega)]^d \subseteq [H^1(\Omega)]^d \hspace{2mm}\textbf{u} = \boldsymbol{\bar{u}} \hspace{2mm} \rm{on} \hspace{2mm} \Gamma_u \} $$
$$\mathscr{C} := \{ c(\textbf{x}) \in [c^o(\Omega) ]^d: c \in [\mathscr{W}(\Omega)]^d \subseteq [H^1(\Omega)]^d \hspace{2mm} c = \bar{c} \hspace{2mm} \rm{on} \hspace{2mm} \Gamma_c \} $$
$$\mathscr{V} := \{ \boldsymbol{v}(\textbf{x}) \in [c^o(\Omega) ]^d: \boldsymbol{v} \in [\mathscr{W}(\Omega)]^d \subseteq [H^1(\Omega)]^d \hspace{2mm} \boldsymbol{v} = \boldsymbol{0} \hspace{2mm} \rm{on} \hspace{2mm} \Gamma_u\} $$
$$\mathscr{P} := \{ p(\textbf{x}) \in [c^o(\Omega) ]^d: p \in [\mathscr{W}(\Omega)]^d \subseteq [H^1(\Omega)]^d \hspace{2mm} p = 0 \hspace{2mm} \rm{on} \hspace{2mm} \Gamma_c \} $$
where the space $\mathscr{W}(\Omega)$ includes linear displacement fields and concentration field. Upon applying the standard Galerkin procedure, the corresponding weak formulation of \eref{eq1} and \eref{eq2} is to find $\boldsymbol{u} \in \boldsymbol{\mathscr{U}}$ and $c \in \mathscr{C}$ such that $\forall \boldsymbol{v} \in \boldsymbol{\mathscr{V}}$ and $\forall p \in \mathscr{P}$:   
 \begin{subequations}
 \begin{equation}
 a(\boldsymbol{u}, \boldsymbol{v}) = \ell(\boldsymbol{v}),
 \end{equation}
 \begin{equation}
  b(c,p) = \ell(p),
 \end{equation}
  \label{weakelasticity}
 \end{subequations}
where 
  \begin{subequations}
\begin{equation}
 a(\boldsymbol{u}, \boldsymbol{v}, c) = \int_\Omega \boldsymbol{\sigma}(\boldsymbol{u}, c) : \boldsymbol{\varepsilon}(\boldsymbol{v}) \hspace{1mm} \rm{d}\Omega,
 \end{equation}
 \begin{equation}
 \ell(\boldsymbol{v}) = {\int_{\Gamma}}_t  \boldsymbol{\hat{t}} \cdot \boldsymbol{v} \hspace{1mm}\rm{d} \Gamma,
 \end{equation}
 \begin{equation}
b(c,p,\boldsymbol{\sigma}_h ) = \int_\Omega \frac{\partial c}{\partial t} p \hspace{1mm}\rm{d} \Omega + \int_\Omega \left( \nabla \textit{p} \right)^{\rm{T}} \nabla c  \hspace{1mm}\rm{d} \Omega - \int_\Omega \left( \nabla \textit{p} \right)^{\rm{T}} c \nabla \boldsymbol{\sigma}_h  \hspace{1mm}\rm{d} \Omega,  
\label{diffusionD}
 \end{equation}
 \begin{equation}
 \ell(p) = \int_\Gamma \boldsymbol{J} \cdot \boldsymbol{n} \hspace{1mm} p\hspace{1mm} \rm{d} \Gamma.
 \end{equation}
 \label{weakdiffusion}
 \end{subequations}
 
  For the temporal discretization, we partition the time interval of interest $\tau$ into $n_{\rm{step}}$ sub-intervals
  %as  $\mathscr{T} = U_{n = 0}^{n_{\rm{step}-1}} \left[t_n,t \right]$ 
  and focus on typical time slab $\left[t_i, t_{i+1}\right]$. To approximate the time derivative of concentration $c$, we apply a first order finite difference scheme as,
  \begin{equation}
      \frac{\partial c}{\partial t} = \frac{c_{i+1} - c_i}{\Delta t}
  \end{equation},
 where $\Delta t = t_{i+1} - t_i > 0$ is the current time increment. We employed the backward Euler time integration scheme and evaluate the discrete set of governing equations at current time step $t_{i+1}$. 
 
FEniCS \cite{AlnaesBlechta2015a} is a very powerful finite element package which translates the weak form of the coupled differential equations into the corresponding finite element solution. The total residual from $F$ used in the FEniCS is given as,
 \begin{equation}
     F = a(\bm{u}, \bm{v}) - \ell(\bm{v}) + b(c,p) - \ell(p).
 \end{equation}
The Newton raphson scheme in the FEniCS can be called as, \emph{solve(F==0, w, bcs)}, where \emph{w} is the solution space and \emph{bcs} are the boundary conditions. Algorithm \ref{procedure}, shows the solution procedure followed in the numerical analysis.

 \begin{algorithm}[htpb]
 \SetAlgoLined
 initialize displacement, concentration fields at time $t_i$- $\bm{u}_i$, $c_i$ \\
 \While {$t_{i+1}<t_{end}$}
 {
  Converged displacement, concentration fields, at time $t_{i}$ - $\vm{u}_{i}$, $\vm{c}_{i}$ \\
  Update strain rates  $\vm{\dot{\varepsilon}}_{i}$, $\vm{\dot{\varepsilon^c}}_{i}$ using backward Euler scheme \\
  \If{ Yield condition satisfied, $f\geq 0$}{Update plastic strain rate $\vm{\dot{\varepsilon^p}}_{i}$}
  Update corresponding stress rate,  $\vm{\dot{\sigma}}_{i}$ and stress field,  $\vm{\sigma}_{i}$ using backward Euler scheme.\\
  Solve for $F=0$ using Newton raphson scheme for displacement and concentration fields, at time $t_{i+1}$ - $\vm{u}_{i+1}$, $\vm{c}_{i+1}$\\
  Update solution for $t_{i}$ to solution at $t_{i+1}$
 } 
 %\nl \bf Pass\;
\caption{Solution algorithm}
\label{procedure}
\end{algorithm}
\section{Problem definition}
\label{S:3}
\begin{figure}[H]
\centering
\subfloat[]{\includegraphics[scale = 0.65]{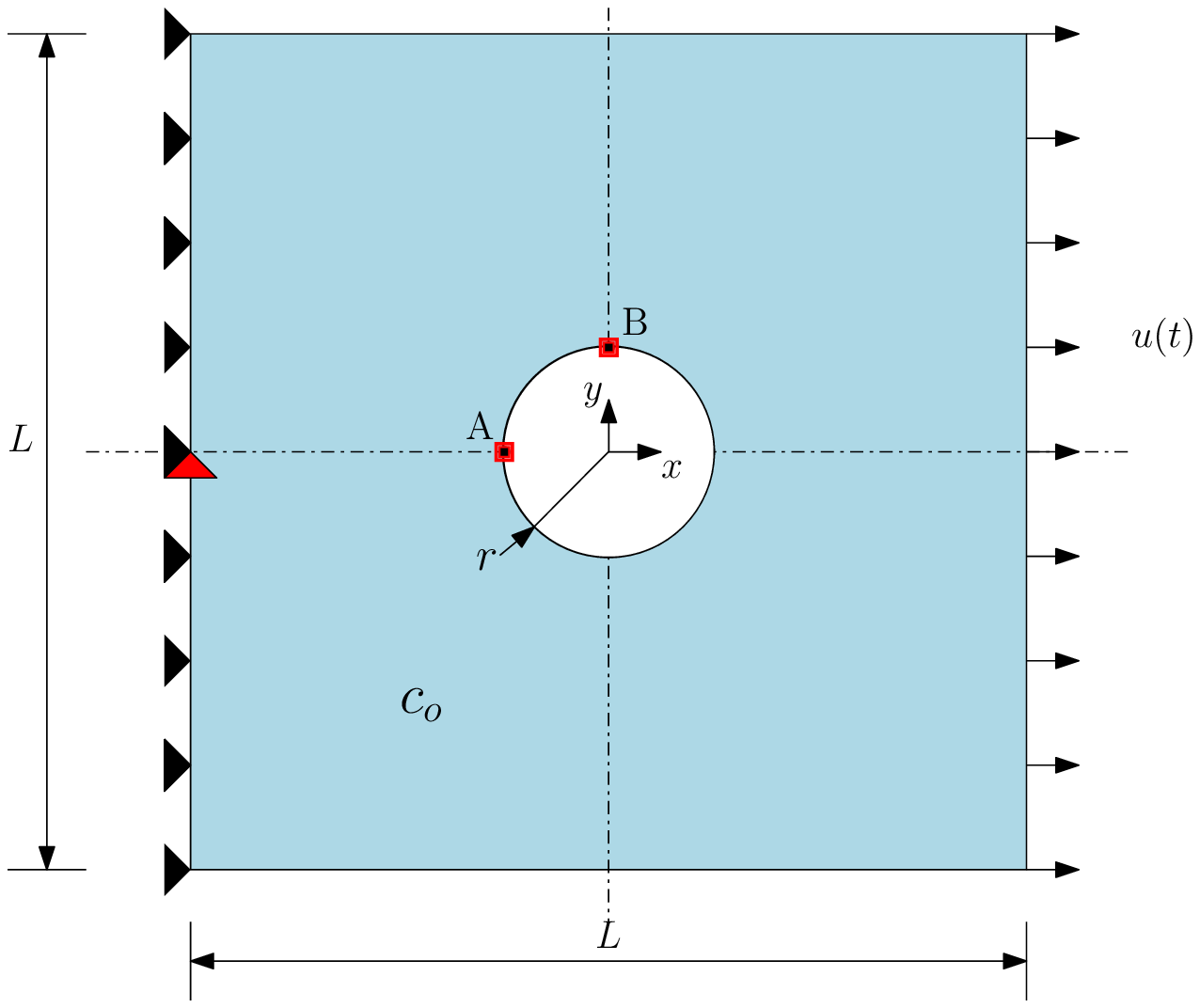}\label{fig:bvp1}} \hspace{10mm}
\subfloat[]{\includegraphics[scale = 0.65]{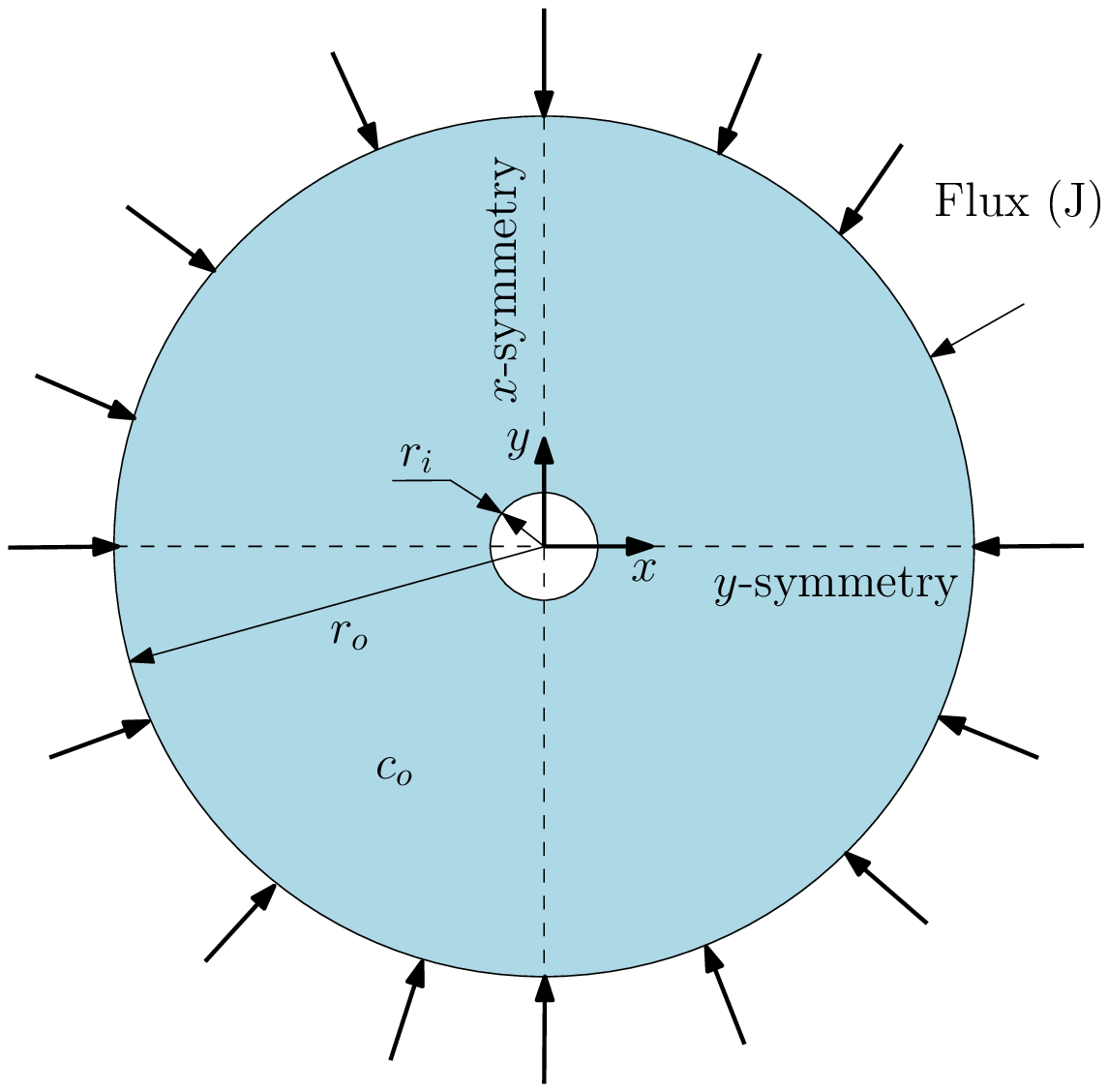}\label{fig:bvp2}}
\caption{(a) Schematic representation of a rectangular domain with a central hole of radius $r$ subjected to uniaxial displacement $u(t)$ at the right edge and fixed at the left edge. The points A, and B are the points of interest for the study. (b) A circular domain of radius $r_o$ with a central void of radius $r_i$ subjected to a radial flux $J$. Both the domains have an initial concentration $C_0$.}
\label{fig:bvp}
\end{figure}
We analyse two boundary value problems in this study: 

\begin{itemize}
    \item a plane strain domain with a circular hole of radius $r$ as shown in \fref{fig:bvp1}. The effects of plasticity and singularity on stress-diffusion interactions is investigated. The radius of the hole $r$ has been varied in order to vary the singularity level on the boundary of the hole. The initial or the reference concentration, $c_0$ is assumed to be zero. The initial and the boundary conditions for the system are $\forall\ t>0$,
$$
u_x ( - L/2, y, t) = 0, \hspace{5mm}
u_y ( - L/2, 0, t) = 0, \hspace{5mm}
u_x (L/2, y,t) = \bar{u}(t), \hspace{5mm}
c( - L/2, y, t) = 1.
$$

\item a plane strain cylindrical domain of outer radius $r_o$ and inner radius $r_i$, see \fref{fig:bvp2}. The initial concentration $c_o$ is assumed to be zero and outer boundary is subjected to constant Flux $J$.   
\end{itemize}

\begin{table}[H]
  \begin{center}
    \caption{Material parameters for Li in graphite system \cite{Swaminathan2016}}
    \label{tab:table2}
    \begin{tabular}{l|l}
      \hline
      Property & Value \\ % <-- added & and content for each column
      \hline
      Young's modulus $(\rm{E})$ & $19.25 \rm{GPa}$\\ % <--
      Poisson's ratio $(\nu)$ & $0.3$\\ % <--
      Diffusivity $(\rm{D})$ & $3.9 \times 10^{-14} \rm{m^2/s}$\\ % <--
      Temperature $(\rm{T})$ & $300 \rm{K}$\\
      Partial molar volume $(\Omega)$ & $4.17 \times 10^{-6} \rm{m^3/mol}$ \\ \hline
     \end{tabular}
  \end{center}
\end{table}

The material parameters used for both the boundary value problem is tabulated in the Table \ref{tab:table2}, which corresponds to the Li in the graphite system \cite{Swaminathan2016}.
\section{Results and Discussions}
\label{S:4}
In this section, the results pertaining to the interactions between stress and concentration will be presented. We analyse problems with increasing complexities. first, we start with (Section \ref{model_validation}) the analysis of the two-dimensional diffusion in pure elastic medium and validation with the analytical solution. Then we validate the current implementation (one-way coupled elasto-plastic simulation) with the commercial available finite element package Abaqus \cite{abaqus2012} and analyzed the effect of two-way coupling. The effect of plastic yielding on the stress-diffusion interaction will be discussed in the Section \ref{plasticity_effect}.  %The effect of the  singularity level will be discussed in the Section \ref{singularity_level}. 
The effect of one-way and two-way for the particle model with and without the plasticity for applied flux is investigated in the Section \ref{section:particle}.

\subsection{Implementation validation with the analytical model and the effect of plasticity }
\label{model_validation} 
The implemented model is validated with the analytical solution available for the pure diffusion. The analytical solution for a circular hole of radius $a$ in an infinite body under plane deformation and tensile forces $p$, the hydrostatic stresses in polar coordinates $(r,\beta)$ are given by \cite{STASHCHUK201214687}:
\begin{equation}
 \sigma_h = \dfrac{(1+\nu)p}{3}\bigg(\dfrac{2R^2}{r^2}cos2\beta - 1\bigg),   
\end{equation}
and the concentration is given by:
\begin{equation}
C(r,\beta) =  C_0e^{\left(\ -k\frac{2(1+\nu)R_0^2p}{3r^2}cos2\beta + C_0Q\right)} e^{\left[ -LambertW \left( C_0e^{\left(-k\frac{2(1+\nu)R_0^2p}{3r^2}cos2\beta + C_0.Q\right)} \right) \right]}
\end{equation}

where $k = \frac{V_H}{RT}, Q = \frac{2\alpha_cV_HE}{9(1-\nu)RT} $, $LambertW(x)$ is the Lambert function.
 
\subsubsection{Diffusion in the pure elastic medium}
\label{diffusion_elastic}
In order to validate the FEniCS implemented model, a rectangular domain with a circular hole with aspect ratio ({L/r}) = 10 \& 20, where L is the length of the domain and $r$ is the radius of the hole, is considered. The initial and the boundary conditions for the system are $\forall\ t>0$,
$$
\sigma_x (-L/2, y, t) = -p, \hspace{5mm}
\sigma_x (L/2, y, t) = p, \hspace{5mm}
c(x, y, t=0) = C_0 \text{ (outside the hole)},
$$
where $p$ is the uni-axial tensile load of 100 MPa magnitude. The boundary of the domain is assumed insulated such that species can not leave the system. The effect of plasticity is neglected and material is assumed to be perfectly elastic for comparison. The material properties of the specimen is tabulated in the Table \ref{tab:table1}.
\begin{table}[H]
  \begin{center}
    \caption{Properties of the steel specimen used for analytical and numerical comparison\cite{STASHCHUK201214687}}
    \label{tab:table1}
    \begin{tabular}{l|l}
      \hline
      Property & Value \\ % <-- added & and content for each column
      \hline
      Young's modulus $(\rm{E})$ & $210 \rm{GPa}$\\ % <--
      Poisson's ratio $(\nu)$ & $0.3$\\ % <--
      Diffusivity $(\rm{D})$ & $1.27 \times 10^{-8} \rm{m^2/s}$\\ % <--
      Temperature $(\rm{T})$ & $300 \rm{K}$\\
      Partial molar volume $(\Omega)$ & $1.96 \times 10^{-6} \rm{m^3/mol}$ \\ \hline
     \end{tabular}
  \end{center}
\end{table}

% \begin{figure}[htbp]
% \centering
% \subfloat[]{\includegraphics[scale = 0.7]
% {./Figures/Hydrostaticstress_analytical_rpt05.pdf}
% \put(-60,50){\color{black}\scriptsize compressive}
% \put(-60,70){\color{black}\scriptsize tensile}
% \put(-65,60){\color{black}\vector(0,-3){15.5}}
% \put(-65,60){\color{black}\vector(0,3){15.5}}
% \label{hydrostatic_analytical}} \hspace{5mm}
% \subfloat[]{
% \includegraphics[scale = 0.7]
% {./Figures/Concentration_analytical_rpt05.pdf}
% \label{concentration_analytical}}
% \put(-150,80){\includegraphics[scale=0.2]{./Figures/point_angle.eps}}
% \caption{Plate with circular hole: (a) hydrostatic stress, $\sigma_h$ and (b) normalized concentration as a function of angle about the center, $\pi=3.14$ \hsk{write unit of hydrostatic stress, for example [MPa]} }
% \label{Figure2}
% \end{figure}

\begin{figure}[H]
\centering
\subfloat[]{\includegraphics[scale = 0.7]
{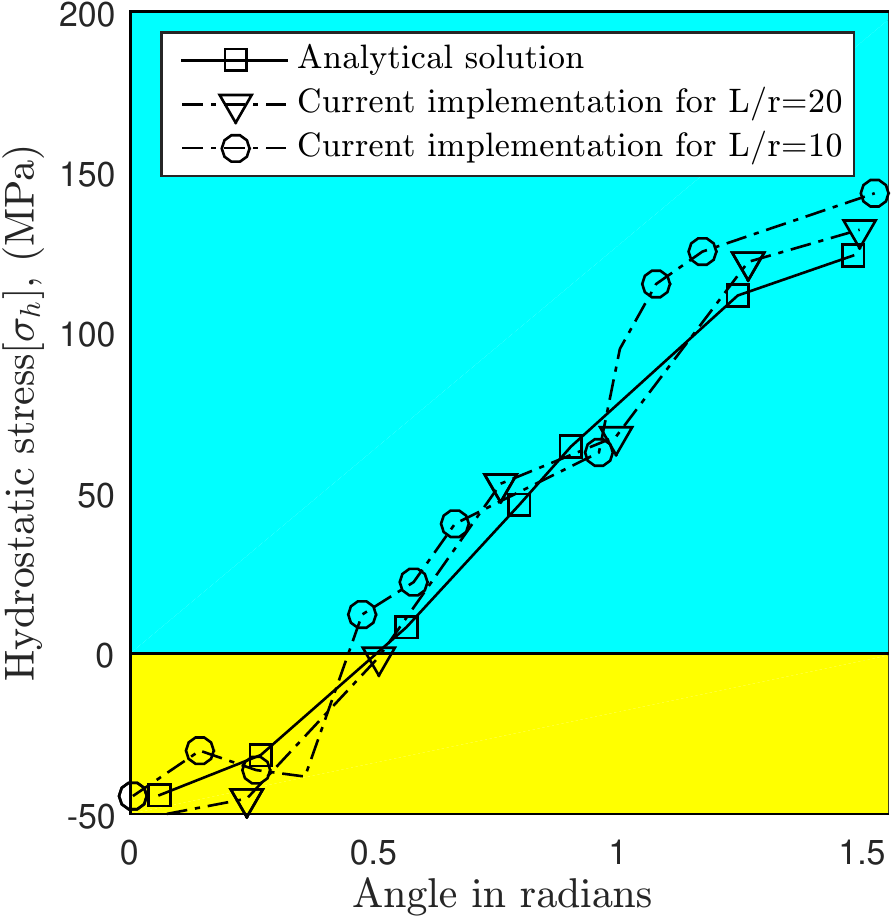}
\put(-60,40){\color{black}\scriptsize compressive}
\put(-60,60){\color{black}\scriptsize tensile}
\put(-65,55){\color{black}\vector(0,-3){15.5}}
\put(-65,55){\color{black}\vector(0,3){15.5}}
\label{hydrostatic_analytical}} \hspace{5mm}
\subfloat[]{
\includegraphics[scale = 0.7]
{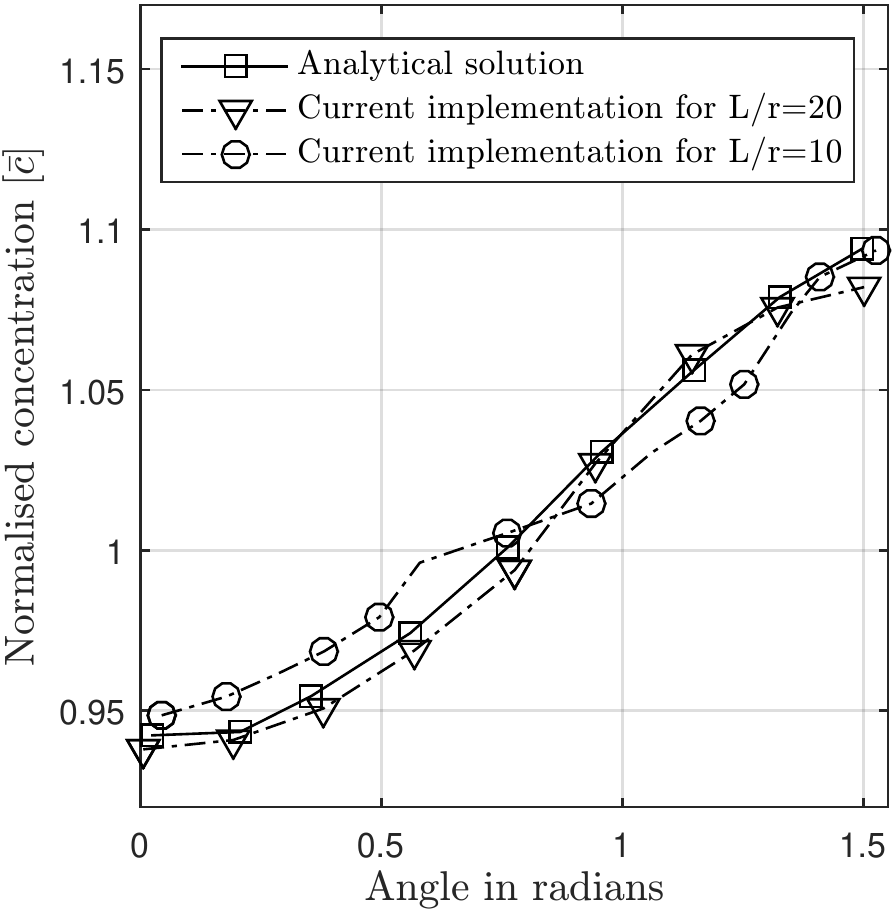}
\label{concentration_analytical}}
\put(-150,80){\includegraphics[scale=0.2]{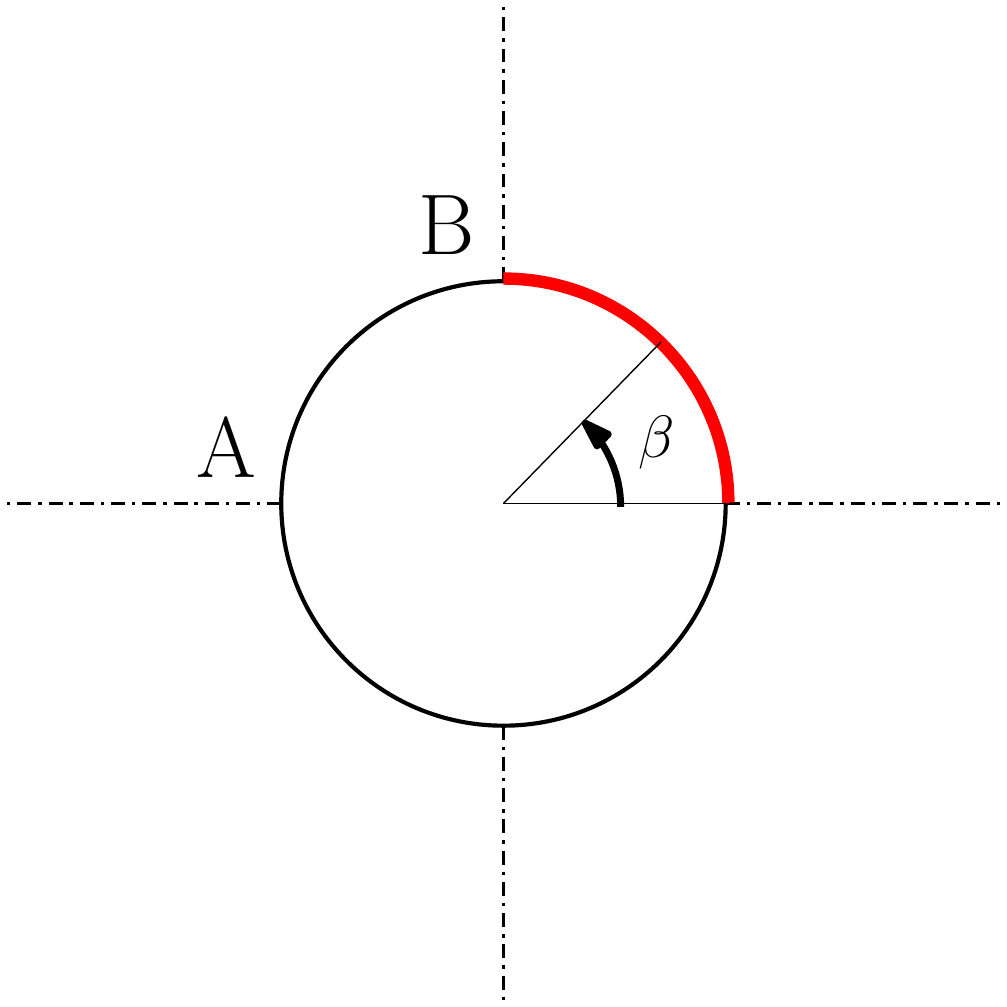}}
\caption{Plate with circular hole: (a) hydrostatic stress, $\sigma_h$ and (b) normalized concentration as a function of angle about the center, $\pi=3.14$ }
\label{Figure2}
\end{figure}

\fref{Figure2} compares the numerical solution for the concentration and the hydrostatic stress with the analytical model as a function of angle about the center. The implemented FEniCS numerical results shows the decent match with the analytical solution as the aspect ratio of the problem domain increases. The state of stress changes from compressive to tensile as the angle increases as shown in \fref{hydrostatic_analytical}. The maximum hydrostatic stress is at $\beta = \pi/2$ whereas the minimum hydrostatic stress is at $\beta = 0$. The distribution of the concentration species follows the distribution of the hydrostatic stress. The tensile sites are capable of holding more diffusion species than the compressive sites and hence the concentration at the tensile sites is more, which is evident from the \fref{concentration_analytical}. 

\begin{figure}[H]
\centering
\subfloat[]{
\includegraphics[scale = 0.7]
{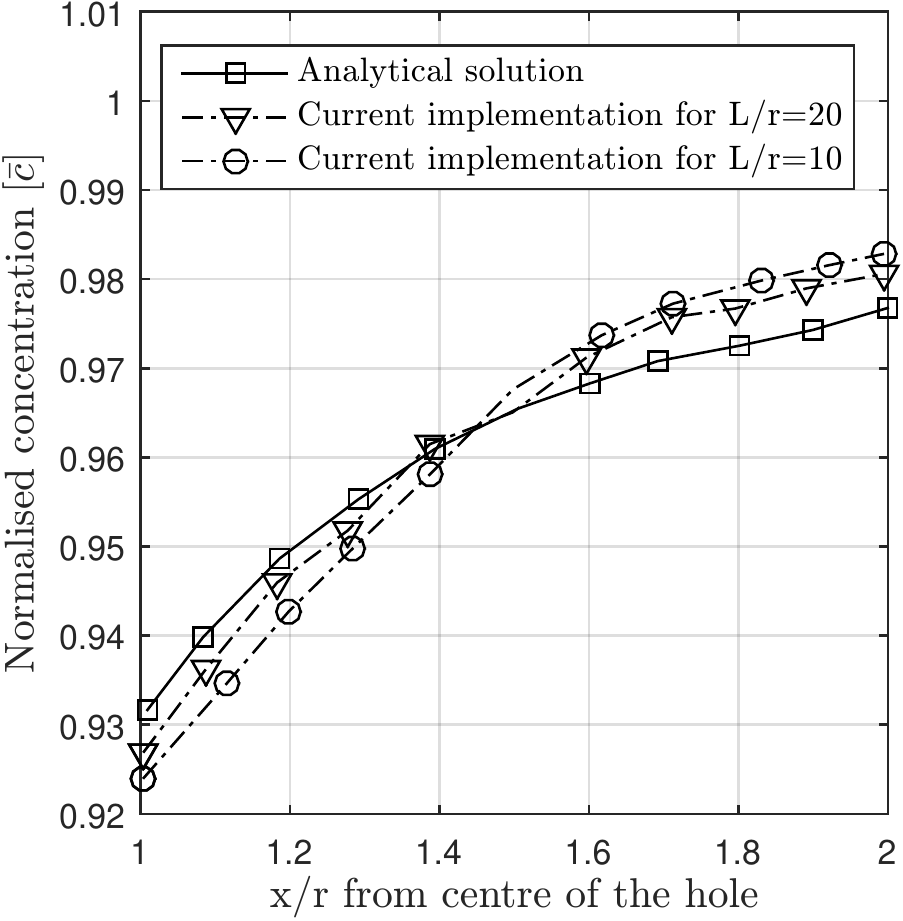}
\label{alonga_n}}\hspace{5mm}
\subfloat[]{\includegraphics[scale = 0.7]
{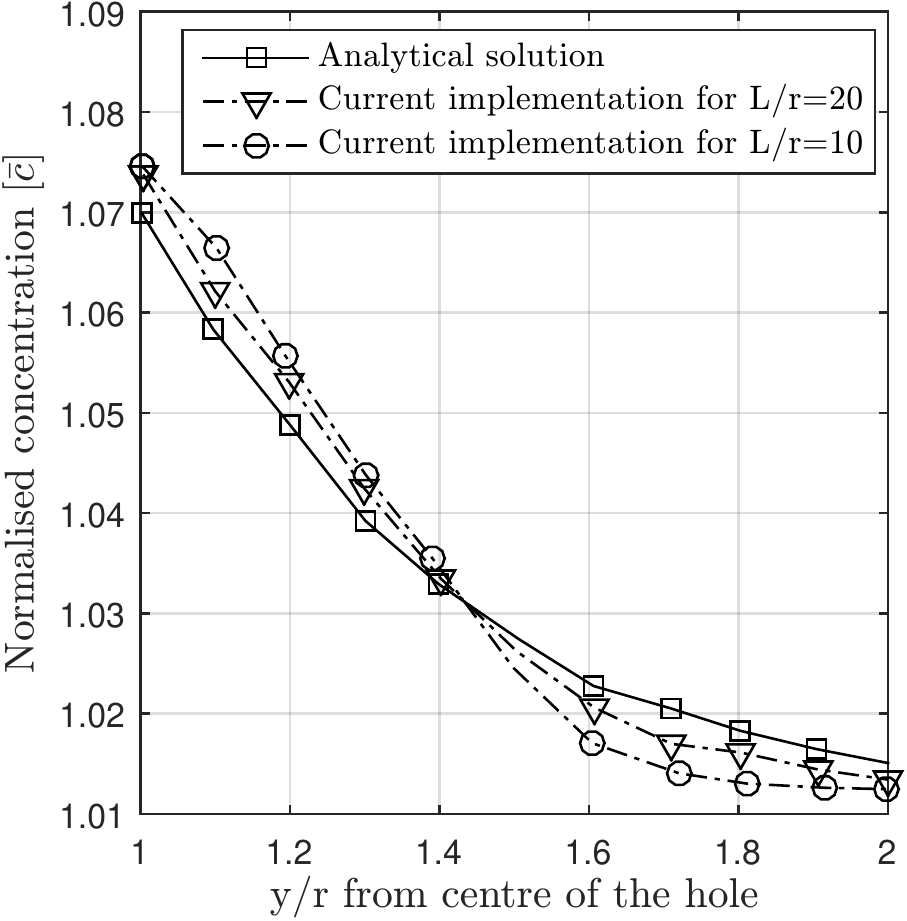}
\label{alongb_n}}
\caption{Plate with a circular hole: normalized concentration along the line (a) $\beta=0$ and (b) $\beta=\pi/2$ }
\label{concentration_analytical_alongr}
\end{figure} 

Now, \fref{concentration_analytical_alongr} compares the concentration of species along the line at $\beta = 0$ and $\beta = \pi/2$. The concentration level shows a decent match with the analytical model for both along the line at $\beta = 0$ and $\beta = \pi/2$. Moreover, The concentration increases as the distance from the compressed region increases as shown in \fref{alonga_n} and decreases as the distance from the tensile region increases as shown in \fref{alongb_n}. Hence singularity level at the localized sites plays a critical role on the concentration build up. 
%We  analyse the effect of singularity in more details in the Section \ref{singularity_level}  

\subsubsection{Effect of plastic yielding on the concentration}

\begin{figure}[H]
\centering
\subfloat[]{\includegraphics[scale = 0.7]
{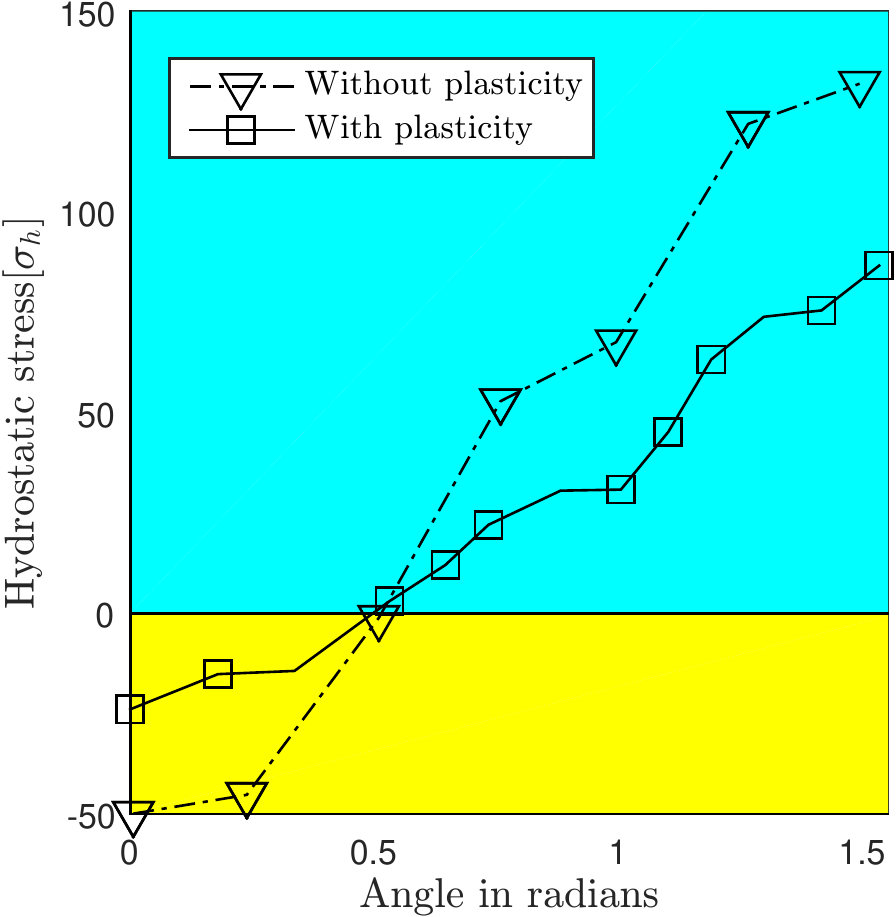}
\put(-60,50){\color{black}\scriptsize compressive}
\put(-60,70){\color{black}\scriptsize tensile}
\put(-65,60){\color{black}\vector(0,-3){15.5}}
\put(-65,60){\color{black}\vector(0,3){15.5}}
\label{hydrostatic_plast_comp}} \hspace{5mm}
\subfloat[]{
\includegraphics[scale = 0.7]
{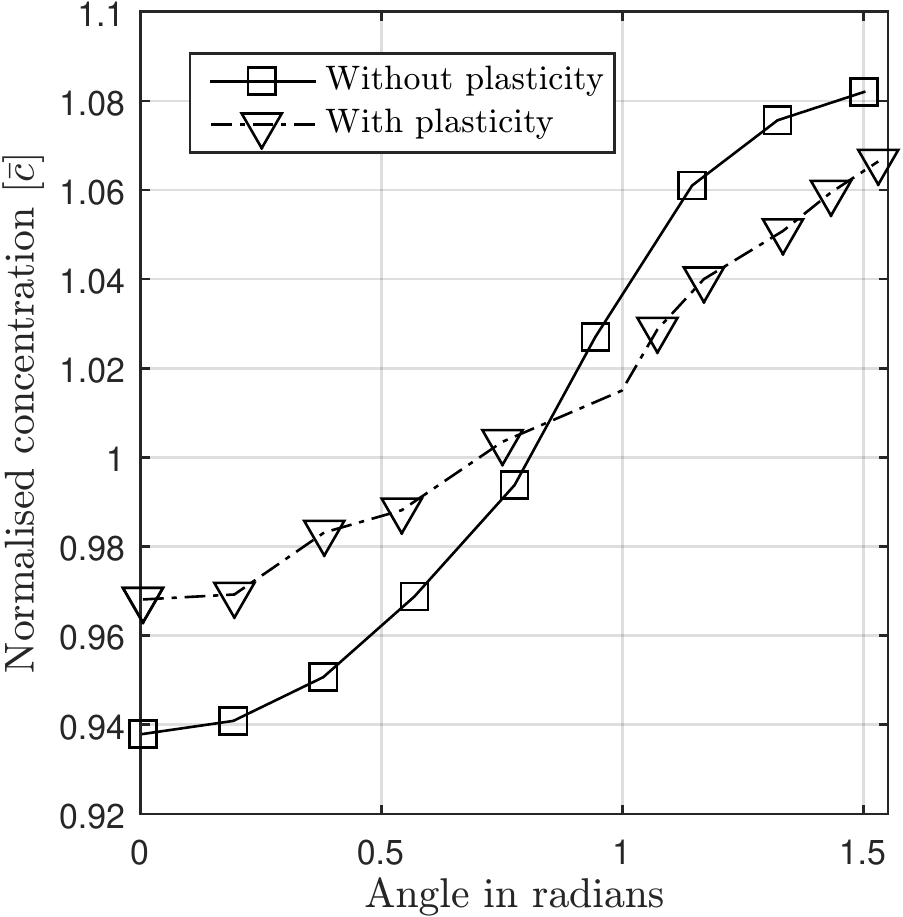}
\label{concentration_plast_comp}}
\put(-150,80){\includegraphics[scale=0.2]{./Figures/point_angle.pdf}}
\caption{Plate with circular hole: (a) hydrostatic stress, $\sigma_h$ and (b) normalized concentration as a function of angle about the center, $\pi=3.14$ }
\label{plasticity_plate_hole}
\end{figure}

We assume a plane strain plate with an elastoplastic medium is subjected to the same boundary conditions as presented in the Section \ref{diffusion_elastic}. \fref{plasticity_plate_hole} shows the hydrostatic stress and the concentration around the hole. The plastic yielding reduces the magnitude of the hydrostatic stress at the compressive as well as the tensile states as shown in \fref{hydrostatic_plast_comp}.

\begin{figure}[H]
\centering
\subfloat[]{
\includegraphics[scale = 0.7]
{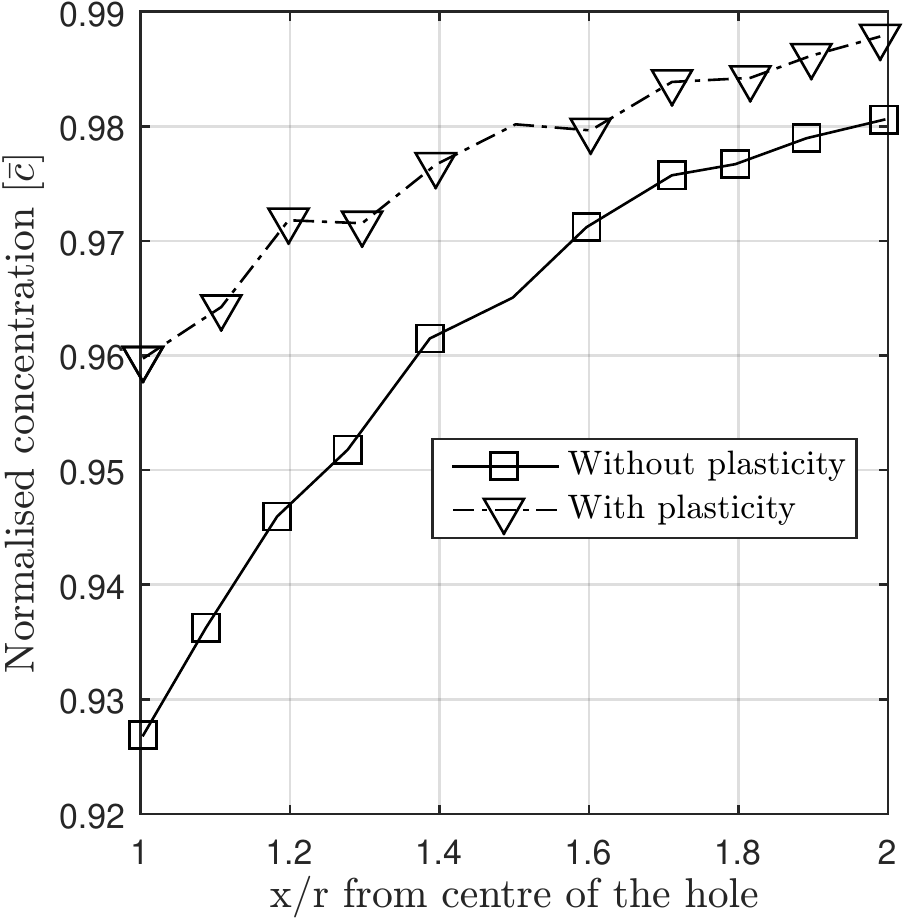}
\label{alonga}}\hspace{5mm}
\subfloat[]{\includegraphics[scale = 0.7]
{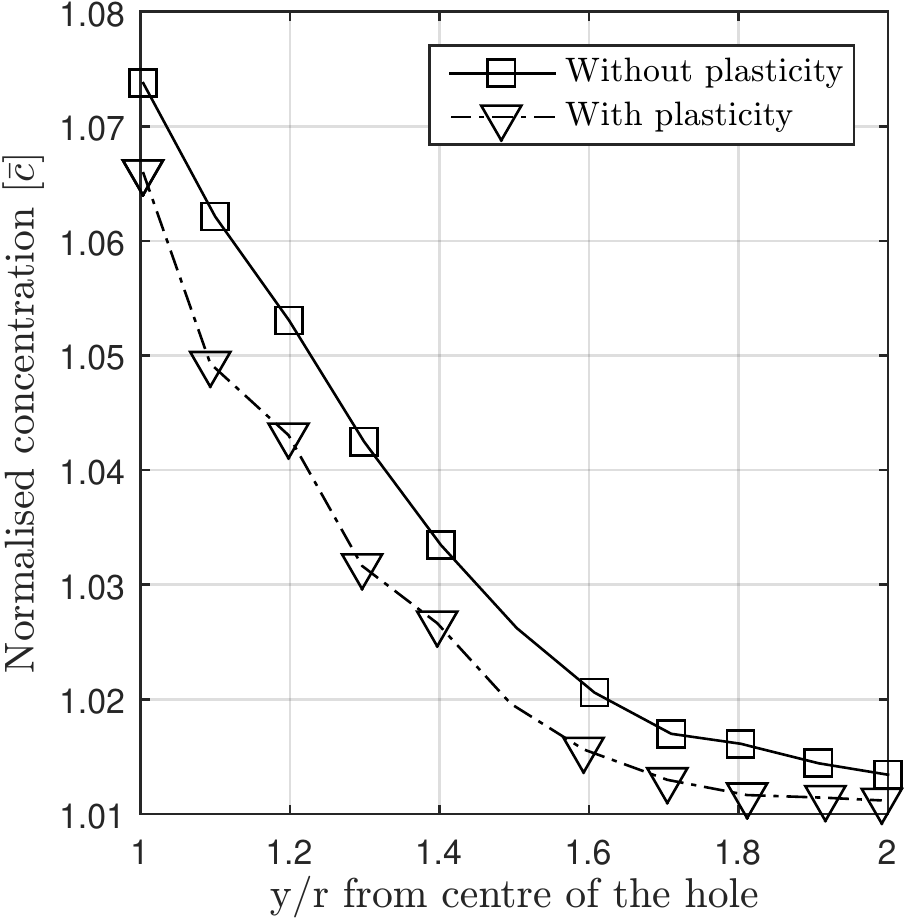}
\label{alongb}}
\caption{Plate with a circular hole: normalized concentration along the line (a) $\beta=0$ and (b) $\beta=\pi/2$ }
\label{concentration_with_wo_plast_alongr}
\end{figure}

The distribution of the concentration species follows the distribution of the hydrostatic stress which has less magnitude than the pure elastic medium at the tensile as well as the compressive sites. The reduced magnitude of tensile stress pulls lesser concentration of diffusing species from the surroundings and the reduced magnitude of compressive stress pushes through lesser concentration species to the surrounding and hence species concentration at the tensile sites is less and more in the compressive sites in elasto-plastic medium than purely elastic medium as shown in \fref{concentration_plast_comp}. \fref{concentration_with_wo_plast_alongr} shows the distribution of the concentration along line $\beta = 0$ and $\beta = \pi/2$ with and without plasticity effect. The concentration level increases as the distance from the singular region increases for elastic as well as elasto-plastic medium for the line $\beta = 0$. The line $\beta = 0$ is in compressive zone and hence the concentration level is more for the elasto-plastic medium than pure elastic medium as shown in \fref{alonga}. The concentration level along $\beta = \pi/2$ decreases for both elastic as well as elastoplastic medium as the distance from the singular region increases but the magnitude of the concentration level is less for the elastoplastic medium as region $\beta  = \pi/2$ is in tensile state of stress as shown \fref{alongb}.

% \begin{figure}[ht!]
% \centering
% \subfloat[]{
% \includegraphics[scale = 0.7]{./Figures/one_way_coupling_concentration_A_Abaquscomp.png}}
% \subfloat[]{
% \includegraphics[scale = 0.7]{./Figures/one_way_coupling_concentration_B_Abaquscomp.png}
% }
% \caption{One way coupling - Normalised concentration vs Normalised time at point (a) A (b) B for a plate with a hole }
% \end{figure}

% \begin{figure}[H]
% \centering
% \subfloat[ ]{
%   \includegraphics[scale = 0.4]{./Figures/ps_t1_2.PNG}
%  } 
%  \subfloat[ ]{
%  \includegraphics[scale = 0.4]{./Figures/ps_t2_2.PNG}
%  } 
% \subfloat[ ]{
%  \includegraphics[scale = 0.4]{./Figures/ps_t3__2_with_legend.PNG}
%  } 
% \caption{Two way coupling - Plastic strain contours at normalized time (a)  0.1 (b) 0.5 (c) 0.95 for a plate with a hole }
% \label{ps_contours}
% \end{figure}

\begin{figure}[!htbp]
 \centering
 \subfloat[]{
 \includegraphics[scale = 0.6]
 {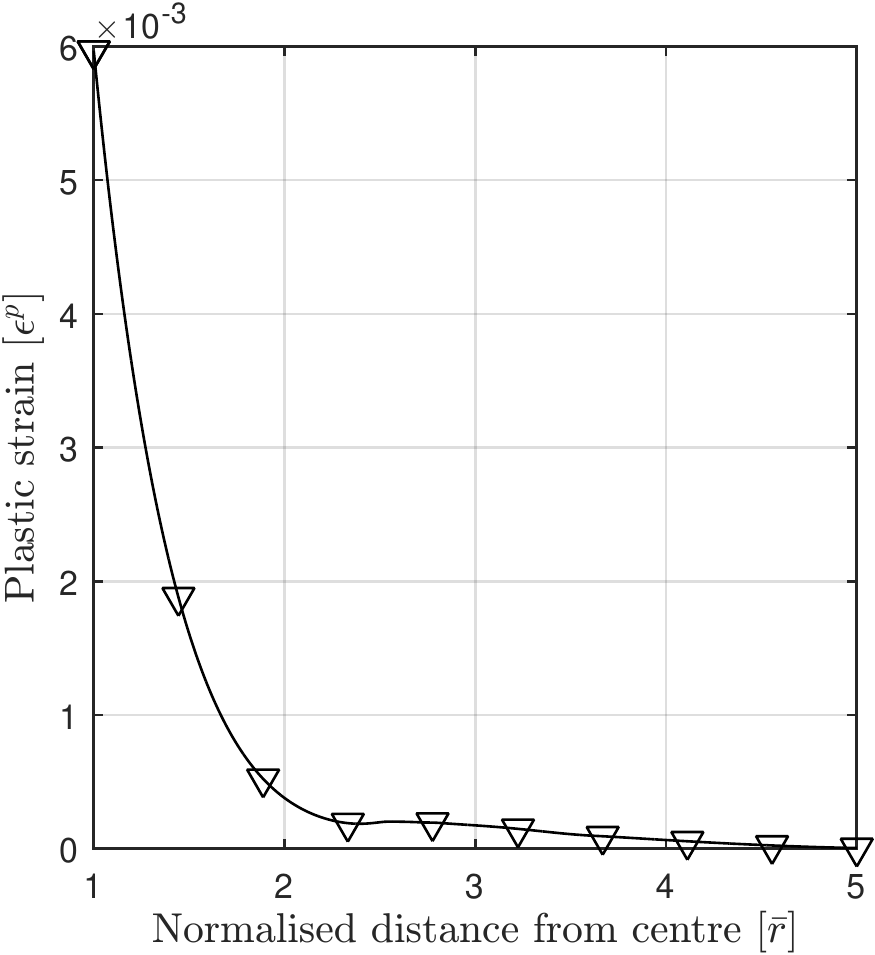}
 \label{psalong0}}\hspace{2mm}
 \subfloat[]{\includegraphics[scale = 0.6]
 {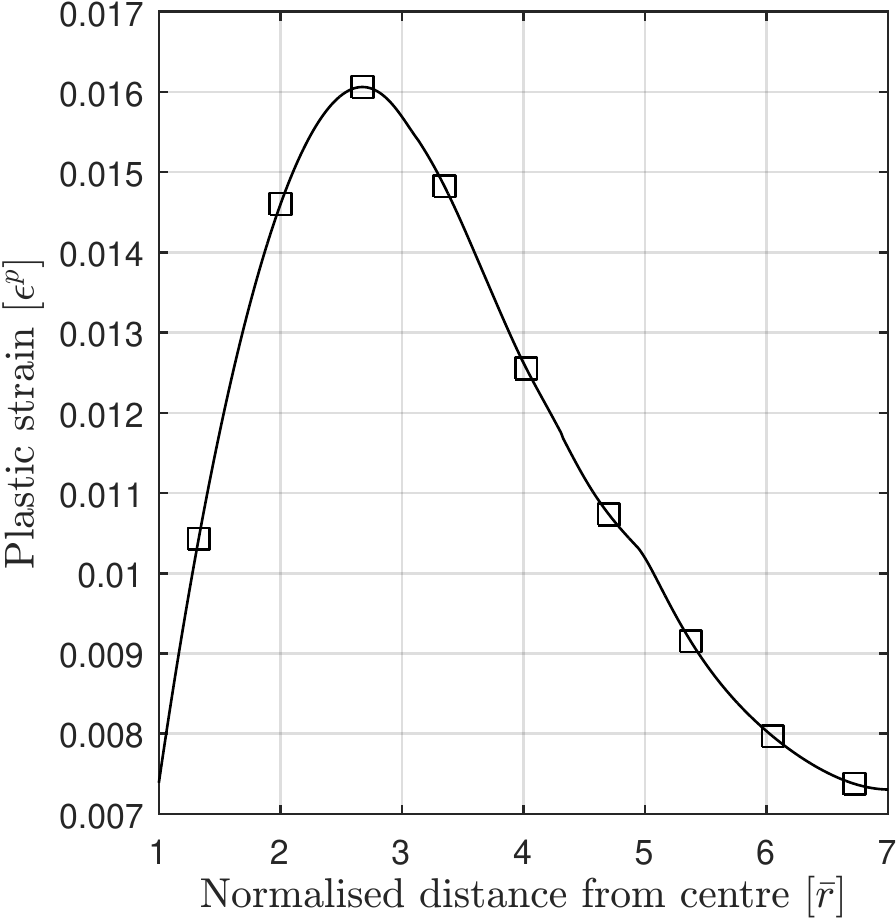}
 \label{psalong90}}\hspace{2mm}
 \subfloat[]{\includegraphics[scale = 0.6]
 {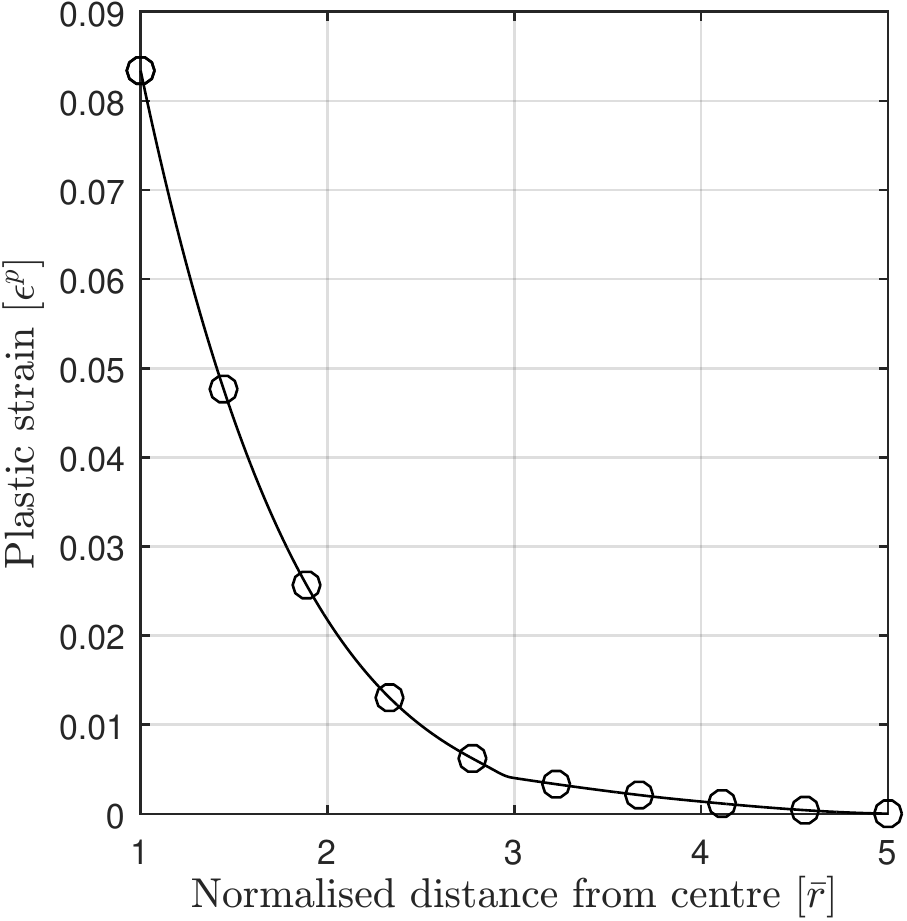}
 \label{psalong45}}
 \caption{Plate with a circular hole: Plastic strain along the line (a) $\beta=0$, (b) $\beta=\pi/4$ and (c) $\beta=\pi/2$}
 \label{plastic_strain_along_r}
\end{figure}

The introduction of the diffusing species in or out of the media affects the magnitude and evolution of the plastic yielding at tensile and compressive region respectively. \fref{plastic_strain_along_r} shows the equivalent plastic strain at normalized time 0.7 along $\beta = 0, \pi/4$ and $\pi/2$. The line  along $\beta = 0$ starts with the compressive zone which implies a reduction of concentration of diffusive species which reduces the stress concentration and hence shows lesser plastic yielding. The line  along $\beta = 0$ starts with the compressive zone which implies a accumulation of concentration of diffusive species which increases the tensile stress and hence shows higher plastic yielding. Away from the singularity the effect of plastic yielding decreases with distance.

% \begin{figure}[H]
% \centering
% \includegraphics[scale = 0.7]{./Figures/platewhole_pstrain_alongr_varyingangle.png}
% \caption{Plate with a circular hole: Plastic strain along the line $\beta=0$, $\beta=\pi/2$ and $\beta=\pi/4$  } 
% \end{figure}

% \begin{figure}[htbp]
% \centering
% \subfloat[]{
% \includegraphics[scale = 0.7]
% {./Figures/one_way_coupling_conc_at_A_B.pdf}}\hspace{5mm}
% \subfloat[]{\includegraphics[scale = 0.7]
% {./Figures/two_way_coupling_conc_at_A_B.pdf}}
% \caption{Normalised concentration vs Normalised time at point A and B for a plate with a hole (a) one-way (b) two-way coupling.}
% \label{one_way_two_way}
% \end{figure}

\subsection{Analysis of boundary value problem a, see \fref{fig:bvp1} }
\label{plasticity_effect}

\begin{figure}[H]
\centering
\subfloat[]{\includegraphics[scale = 0.7]{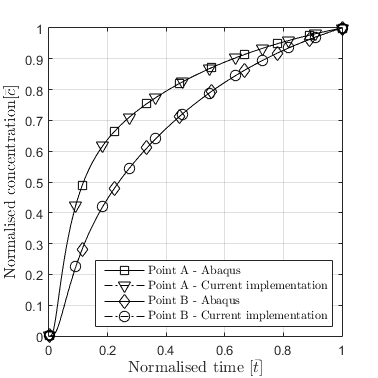}}
\caption{Plate with a hole: normalized concentration with respect to normalized time at points A and B for one-way coupled system} 
\label{fig:abaqus_vs_FEniCS_oneway}
\end{figure}

\begin{figure}[H]
\centering
\subfloat[]{
\includegraphics[scale = 0.7]{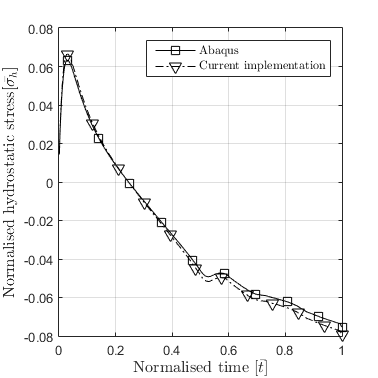}}
\subfloat[]{
\includegraphics[scale = 0.7]{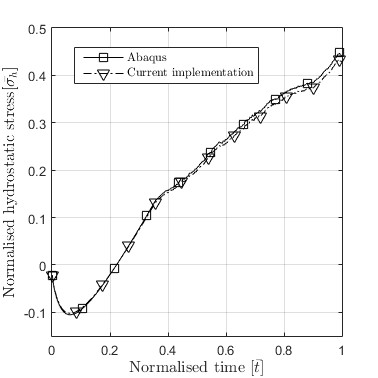}
}
\caption{Plate with a hole: evolution of normalized hydrostatic stress, $\sigma_h$ with respect to the normalized time at point (a) A (b) B for for one-way coupled system.}
\label{fig:Abaqus_vs_FEniCS_stress}
\end{figure}

Before analyzing the problem in detail, we have verified our FEniCS implementation with Abaqus. In Abaqus, we have used coupled-thermal analysis which is one-way coupled analysis in our context. The analysis is performed for the elastoplastic material. \fref{fig:abaqus_vs_FEniCS_oneway} shows the concentration at points A and B with respect to the normalized time for one-way coupled problem. The point A is nearer than the point B from the left boundary and hence the concentration is higher at point A at any point of time. However, both points reaches steady state concentration level at the end of the simulation which is consistent with the work of Natarajan {\it et. al.,}\cite{Natarajan2016}. This is due to the fact that the effect of stresses in not considered in one-way coupling. Our implementation results shows excellent agreement with the Abaqus results, see \fref{fig:abaqus_vs_FEniCS_oneway}. \fref{fig:Abaqus_vs_FEniCS_stress} also compares the hydrostatic stress evolution at points A and B with respect to the normalized time. The implemented model is able to capture tensile to compressive state at point A and compressive to tensile state at point B which is also evident from the Abaqus results.

\begin{figure}[H]
\centering
\subfloat[]{
\includegraphics[scale = 0.7]
{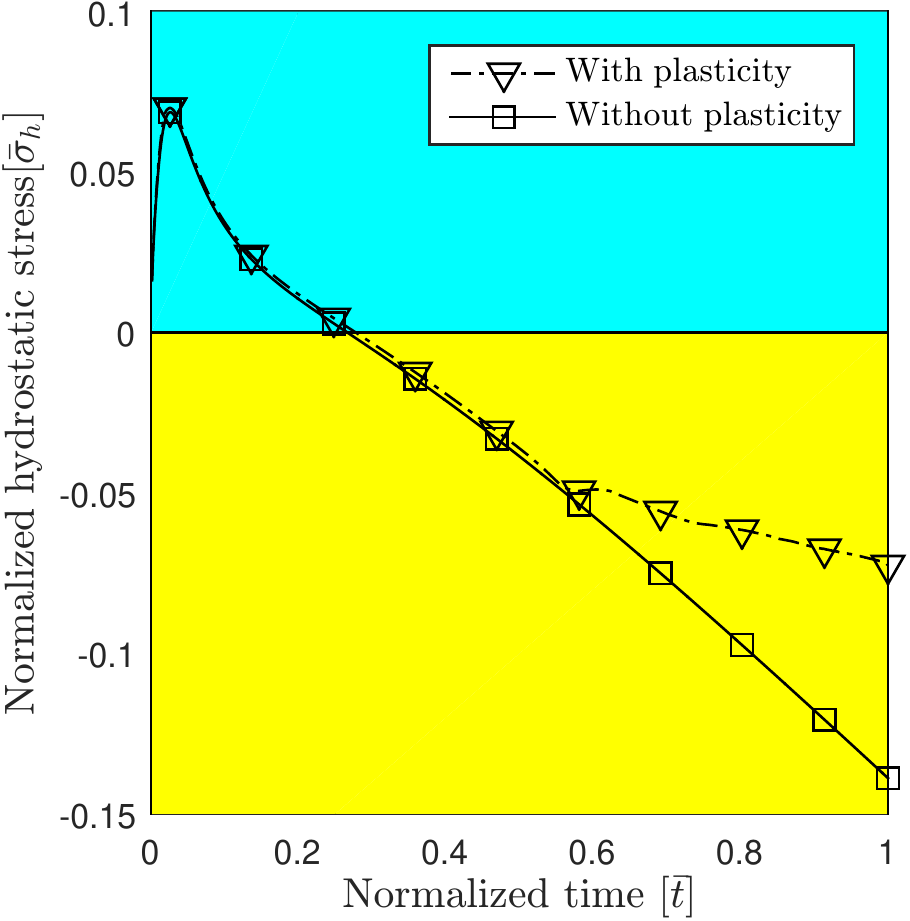}
\label{hydrostatic_stress_plasticitya}} \hspace{5mm}
\subfloat[]{\includegraphics[scale = 0.7]
{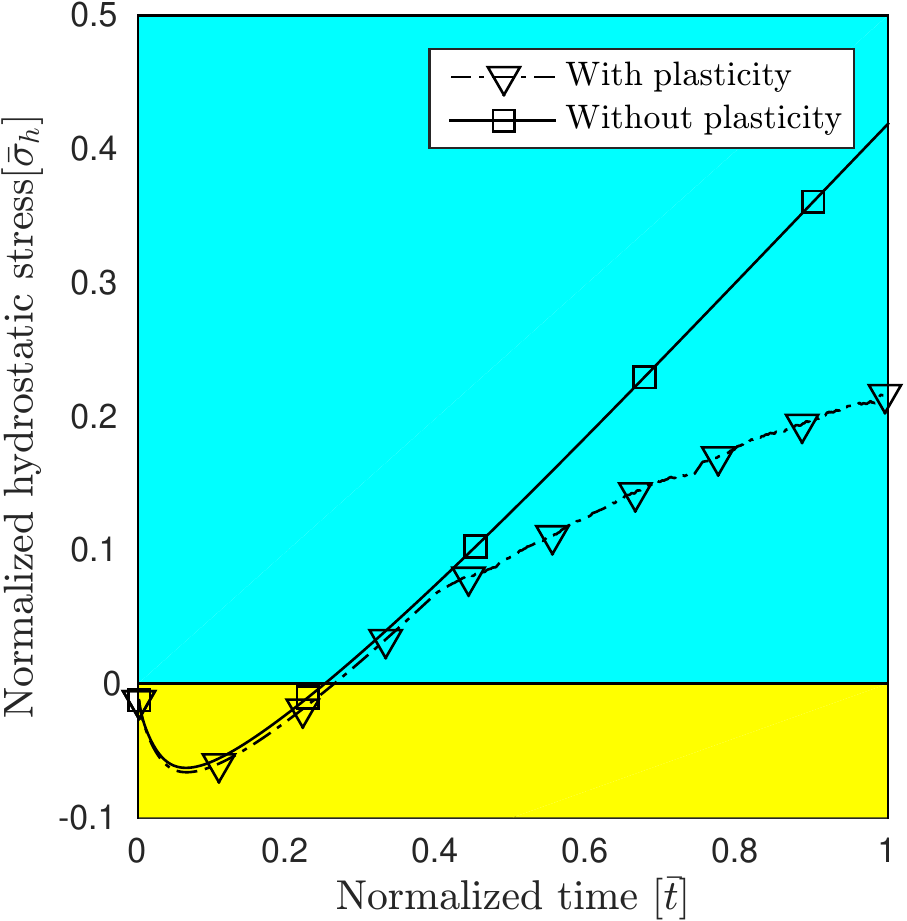}
\label{hydrostatic_stress_plasticityb}}
\put(-320,30){\includegraphics[scale=0.2]{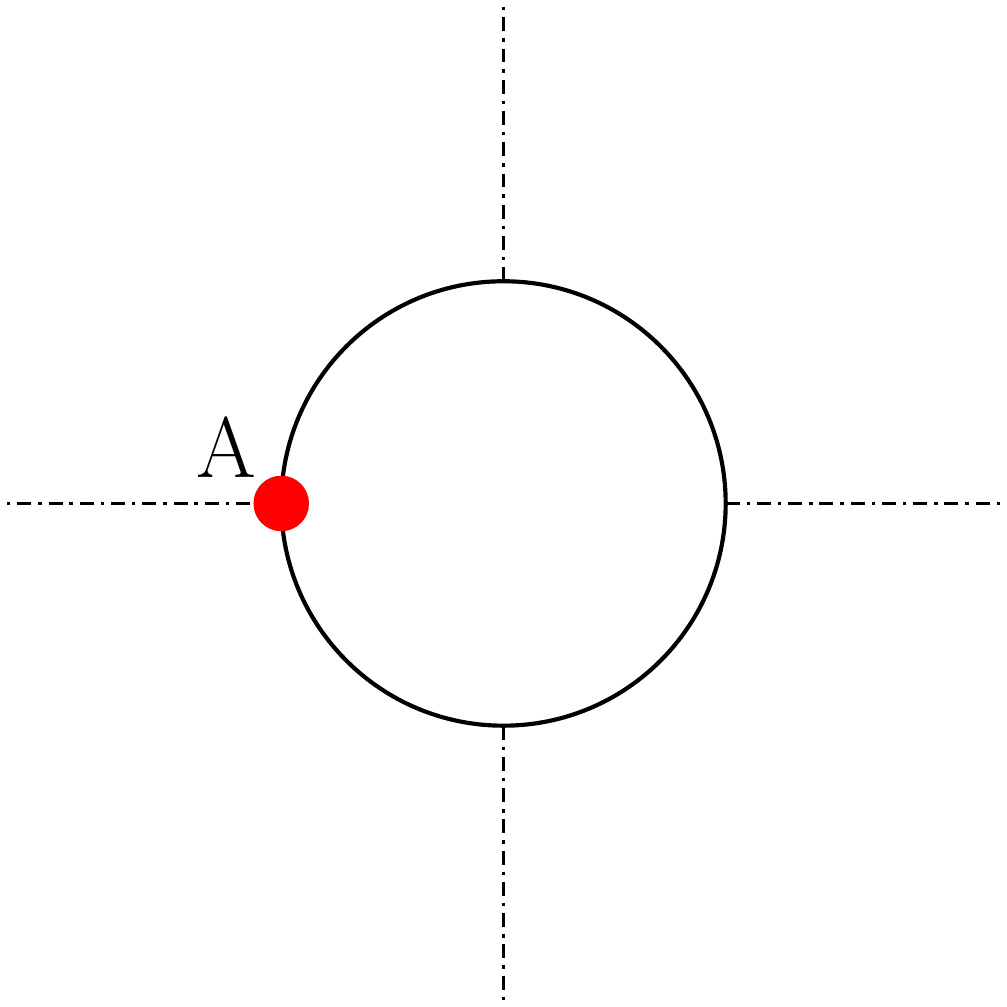}}
\put(-150,70){\includegraphics[scale=0.2]{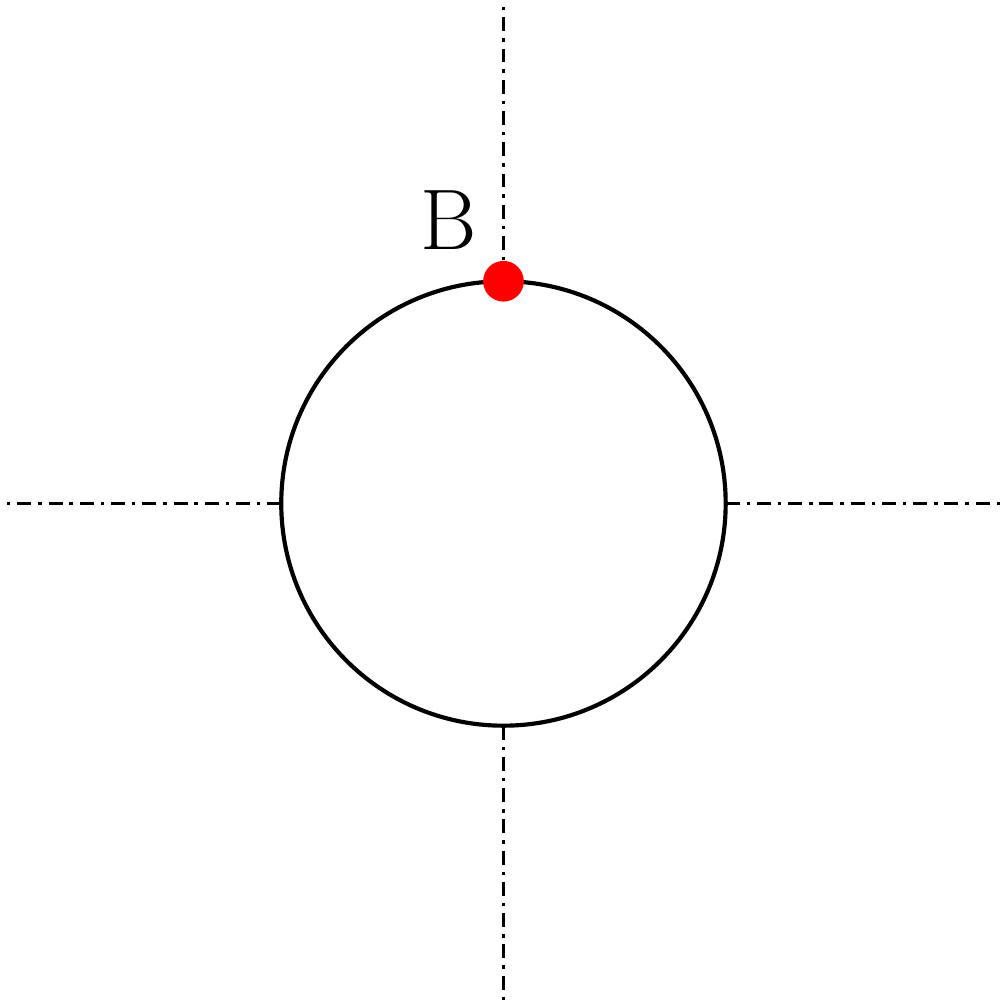}}
\put(-255,110){\color{black}\scriptsize compressive}
\put(-255,125){\color{black}\scriptsize tensile}
\put(-260,119){\color{black}\vector(0,-3){15.5}}
\put(-260,119){\color{black}\vector(0,3){15.5}}
\put(-60,40){\color{black}\scriptsize compressive}
\put(-60,58){\color{black}\scriptsize tensile}
\put(-65,50){\color{black}\vector(0,-3){15.5}}
\put(-65,50){\color{black}\vector(0,3){15.5}}

\caption{Two way coupling - Normalized hydrostatic stress, $\bar{\sigma_h}$ as a function of normalized time at point (a)A (b) B for with and without plasticity.}
\label{hydrostatic_stress_plasticity}
\end{figure}

\begin{figure}[H]
\centering
\subfloat[]{
\includegraphics[scale = 0.7]
{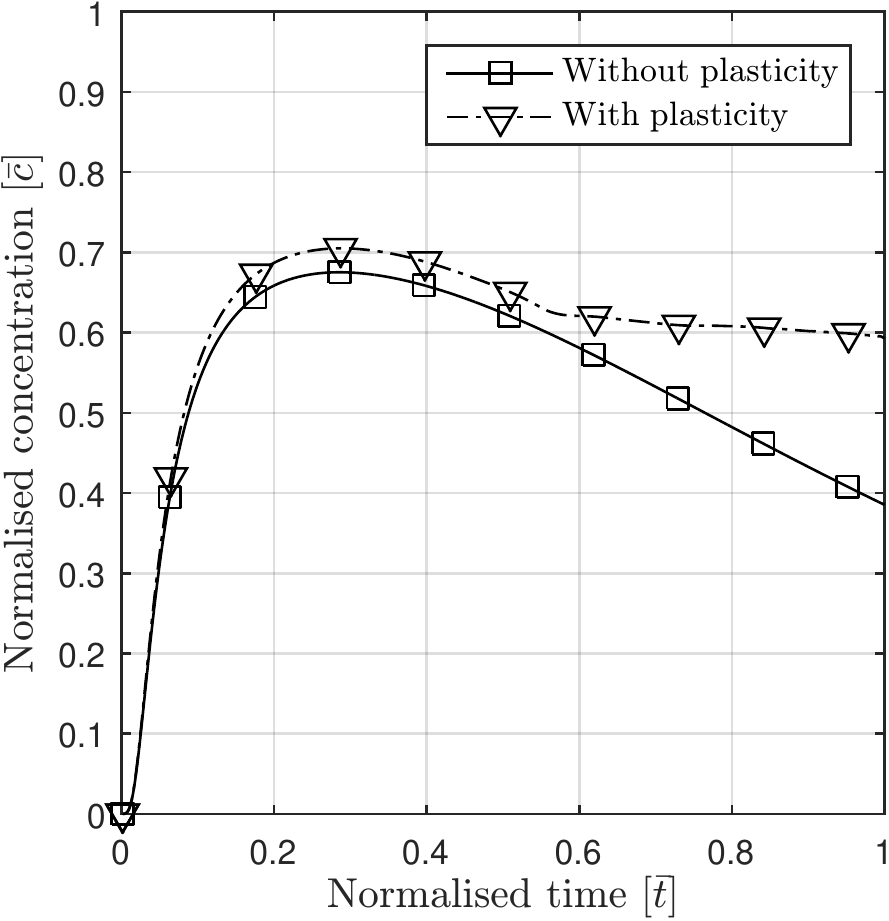}
\label{concentration_plasticitya}}\hspace{5mm}
\subfloat[]{\includegraphics[scale = 0.7]
{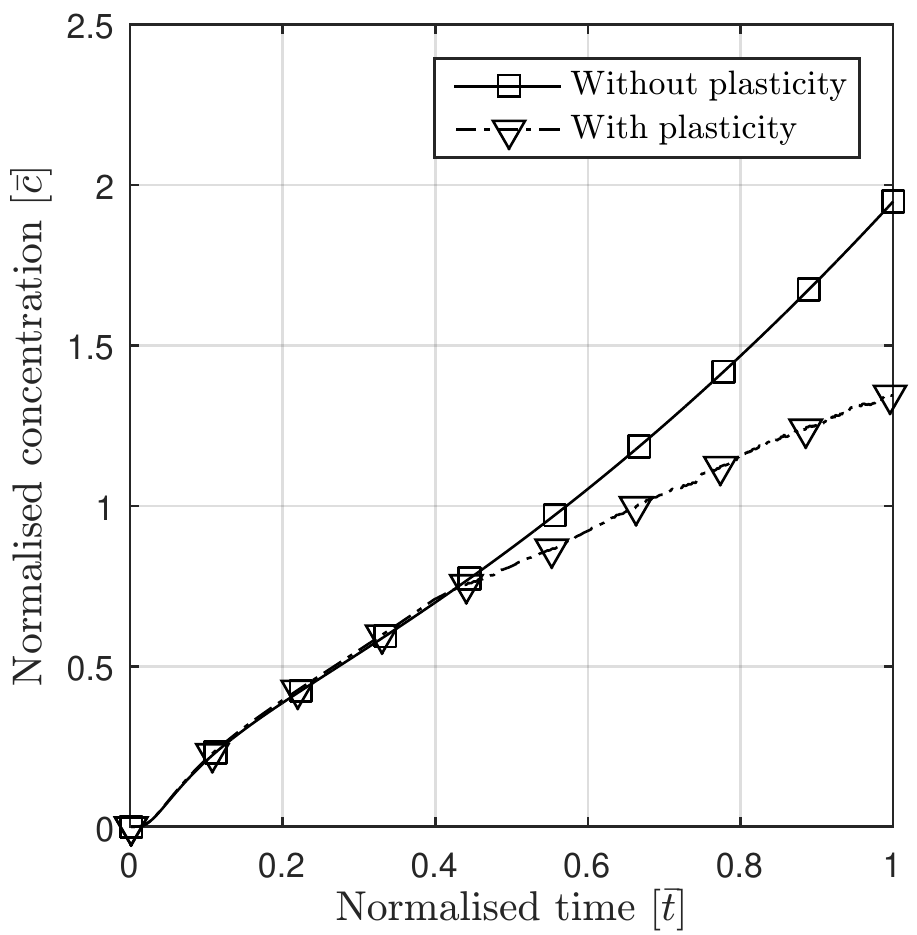}
\label{concentration_plasticityb}}
\put(-320,50){\includegraphics[scale=0.2]{./Figures/pointA.pdf}}
\put(-150,100){\includegraphics[scale=0.2]{./Figures/pointB.pdf}}
\caption{Two way coupling- Normalized concentration vs Normalized time at point (a) A (b) B for a plate with a hole with and without plasticity}
\label{concentration_plasticity_a}
\end{figure}

In order to illustrate the effect of plasticity on the stress-diffusion interactions, we compare the results of pure elastic and elastoplastic material and two-way coupled model. \fref{hydrostatic_stress_plasticity} shows the hydrostatic stress evolution at different points as a function of normalized time for pure elastic as well as elastoplastic material. The state of stress at a point A is initially tensile but changes to compressive due to continuous pulling and concentration evolution in the domain. Moreover, the state of stress changes from compressive to tensile at point B. The plastic yielding significantly reduces the stress level in the tensile site (point B) and makes it less compressive in the compressive site (point A). The level of concentration in the domain depends on the gradient of concentration as well as the localized state of stress. Tensile sites are capable to hold more concentration whereas compressed sites tries to push the concentration to near by sites. \fref{concentration_plasticitya} shows the concentration evolution at the point A. Initially the state of stress at this point is tensile and gradient of concentration is also present, the concentration of species increases with respect to time. But as the state of stress changes from tensile to compressive (at normalized time 0.3 please refer \fref{hydrostatic_stress_plasticitya}),  the point A is no longer able to take/hold the concentration of the species and hence magnitude of concentration of the species starts decreasing. However in an elastoplastic medium the compressive stress is reduced due to plastic yielding which reflects on the higher concentration compared to pure-elastic medium. In contrast at point B, the level of concentration increases monotonically as shown in \fref{concentration_plasticityb}. The initial increase in the concentration level is due to the gradient of the concentration and later gradient of hydrostatic term predominates which is responsible for the increase of concentration level. Once again, however we notice that the tensile stress reduction due to plastic yielding reduces concentration of the diffusing species at point B in an elastoplastic medium. Hence it can be concluded that the plastic yielding in the system plays a critical role in determining the steady state behaviour.

\begin{figure}[H]
\centering
\subfloat[ ]{
  \includegraphics[scale = 0.38]{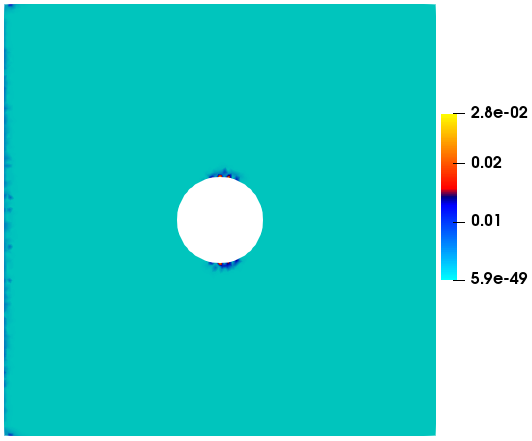}
} 
 \subfloat[ ]{
 \includegraphics[scale = 0.38]{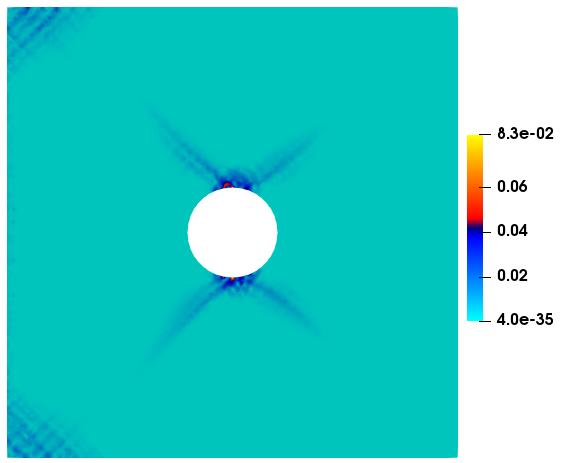}
 } 
\subfloat[ ]{
 \includegraphics[scale = 0.38]{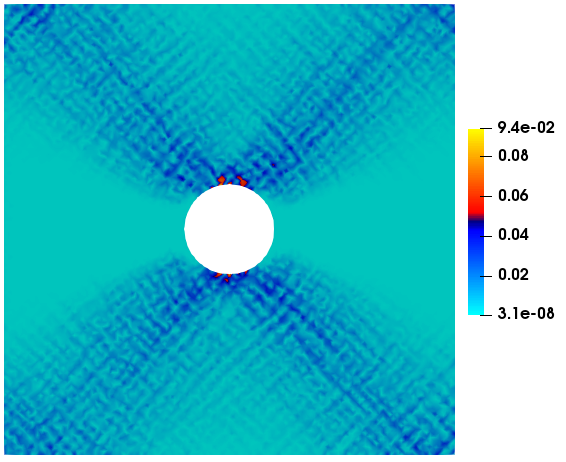}
 } 
\caption{One-way coupling - Plastic strain contours at normalized time (a)  0.30 (b) 0.4 (c) 0.7 }
\label{fig:plasticstrain_oneway}
\end{figure}

\begin{figure}[H]
\centering
\subfloat[ ]{
  \includegraphics[scale = 0.38]{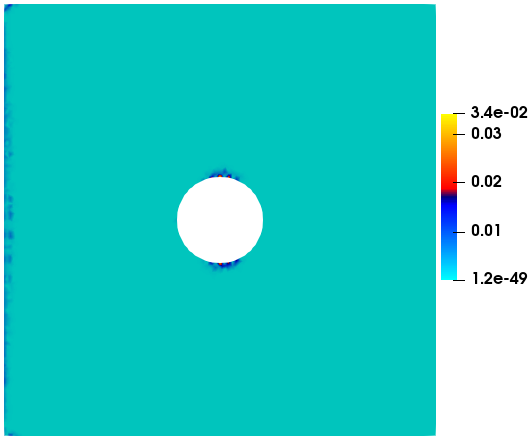}} 
 \subfloat[ ]{
 \includegraphics[scale = 0.38]{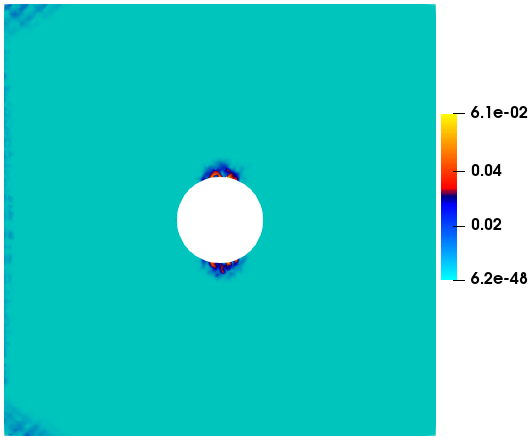}
 } 
\subfloat[ ]{
 \includegraphics[scale = 0.38]{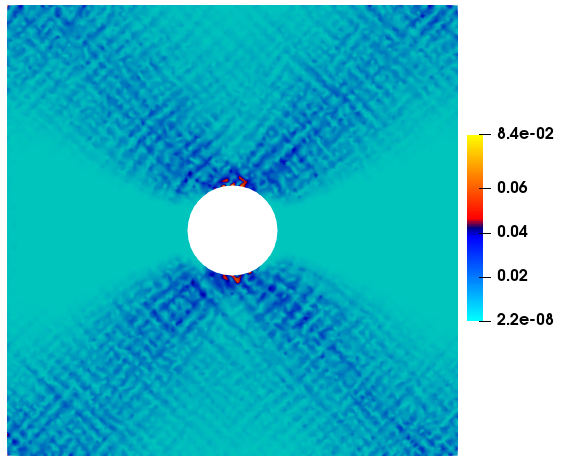}
 } 
\caption{Two-way coupling - Plastic strain contours at normalized time (a)  0.35 (b) 0.4 (c) 0.7}
\label{fig:plasticstrain_twoway}
\end{figure}

% \begin{figure}[H]
% \centering
% \subfloat[]{
% \includegraphics[scale = 0.35]{./Figures/plast_straint_1500_abaqus.png}}
% \subfloat[]{
% \includegraphics[scale = 0.35]{./Figures/plast_straint_1500_fenics.png}
% }
% \caption{Plate with a hole: at t=1.5 sec for one-way coupled system.}
% \label{fig:Abaqus_vs_FEniCS_pstrain}
% \end{figure}

Figures \ref{fig:plasticstrain_oneway} and \ref{fig:plasticstrain_twoway} shows the plastic strain evolution for one-way and two-way coupled system at different normalized time. In the two-way coupled system, there is a local stress relaxation (analogous
to expansion due to temperature rise \cite{Natarajan2016}) and therefore, it leads to a lower stress concentration around the singularity. The equivalent plastic strain due to tension around the hole is lesser in case of two-way coupled system than the one-way coupled system, see Figures \ref{fig:plasticstrain_oneway} and \ref{fig:plasticstrain_twoway}. This provides evidence of the reduced tensile stress concentration and the concentration of the species due to the relaxation effect in case of two-way coupled system. 

\begin{figure}[H]
\hspace{10mm}one-way coupled system \hspace{18mm}two-way coupled system \\
\centering
\subfloat[ ]{
  \includegraphics[scale = 0.45]{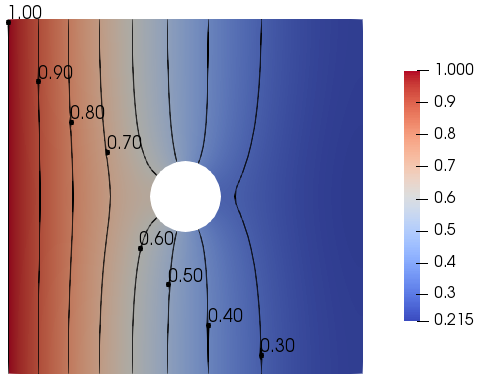}\label{conc_onewaya}} \hspace{5mm}
\subfloat[ ]{
  \includegraphics[scale = 0.45]{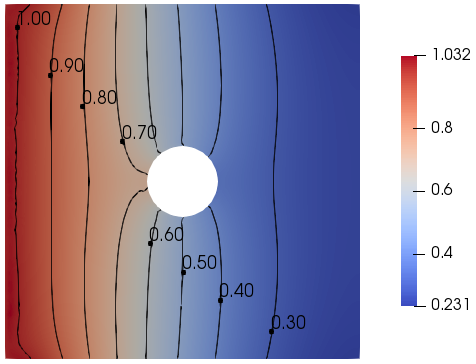}\label{conc_twowaya}}

  \subfloat[ ]{
 \includegraphics[scale = 0.45]{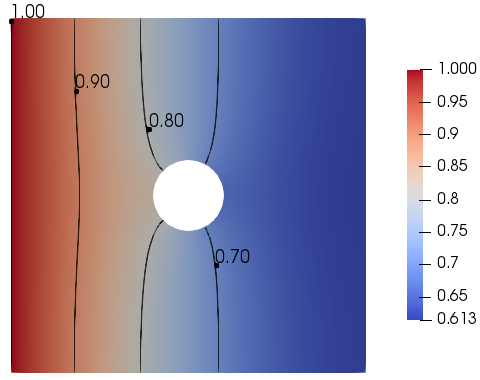}\label{conc_onewayb} } \hspace{5mm}
   \subfloat[ ]{
 \includegraphics[scale = 0.45]{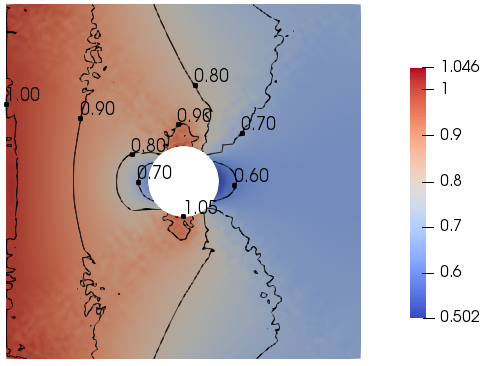}\label{conc_twowayb} }

 \subfloat[ ]{
 \includegraphics[scale = 0.45]{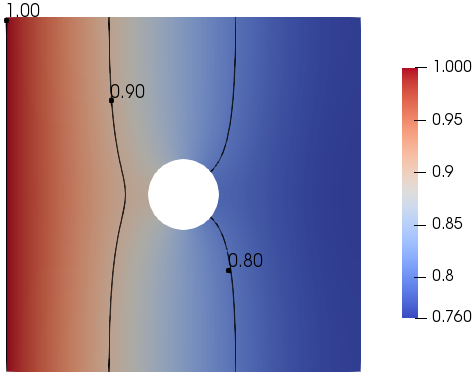}\label{conc_onewayc} } \hspace{5mm}
 \subfloat[ ]{
 \includegraphics[scale = 0.45]{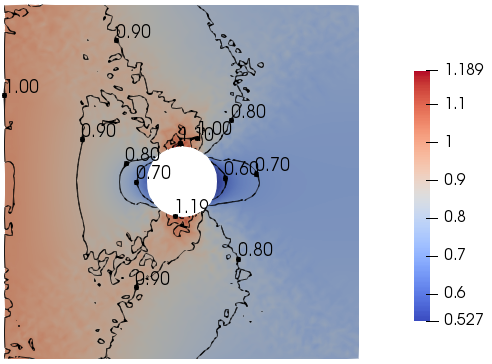} \label{conc_twowayc}
 }
    \caption{Plate with hole: concentration contour at normalized time (a) 0.3 (b) 0.4 and 0.7 for one-way and two-way coupled system.}
    \label{fig:plateHole_concentration_one_vs_two}
\end{figure}

\fref{fig:plateHole_concentration_one_vs_two} compares the concentration contour of one-way and two-way coupled system at different normalized time. The asymmetry of the concentration in the case of one-way coupling is only due to the hole present in the system, see Figures \ref{conc_onewaya},\ref{conc_onewayb} and \ref{conc_onewayc}. Whereas the asymmetry of the concentration is due to both hole and the stresses, see Figures \ref{conc_twowaya}, \ref{conc_twowayb} and \ref{conc_twowayc}. In the case of Two-way coupling, the location of the concentration build up is not the same as compared to the one way coupling. Although the point A is nearer to the point B from the left edge, the concentration build up is more at the point B. This observation is opposite to our earlier observation in one-way coupling. The main reason behind the different build up profile is due to stress induced diffusion effect in the case of two-way coupling. Hence in studying the stress diffusion interactions consideration of the effect of stresses on the singularities is mandatory. The effect is more dominant if the high stress gradient present in the domain as in case of stress singularities due to discontinuities.

\subsection{Analysis of boundary value problem b, see \fref{fig:bvp2}}
\label{section:particle}
% \begin{figure}[H]
% \centering
% \includegraphics[scale = 0.7]{./Figures/particle_model_bvp.PNG}
% \caption{ Schematic representation of the boundary value problem for Lithium particle with a void } 
% \end{figure}
% $$
% u_y(x,0,t) = 0,  u_x(0,y,t) = 0,  J(r=r_o,t) = J_0
% $$
In the recent times, hollow nano-anode structures have attracted attention because of their extra space to accommodate the volume expansion/contraction due to charge and discharge and hence increases the cyclic performance. We attempt to understand the influence a void in a particle on the chemo-mechanical behaviour for elasto-plastic material. 

\begin{figure}[H]
\centering
\subfloat[]{\includegraphics[scale = 0.7]{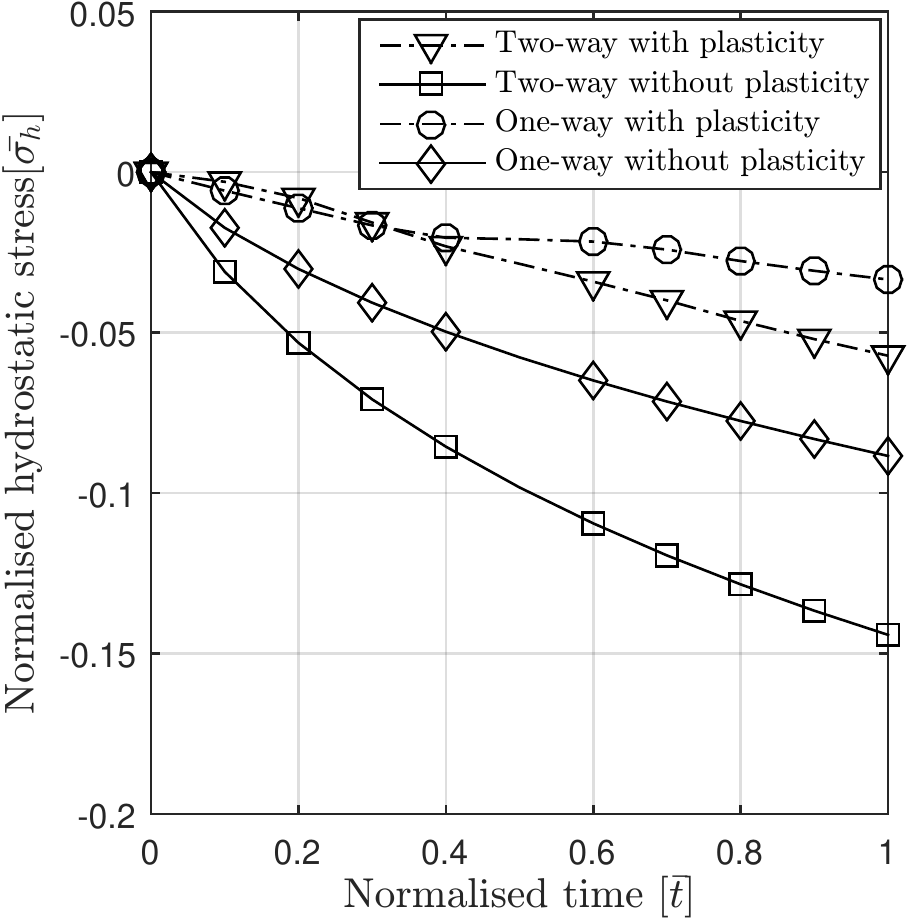}\label{fig:particle_stress_outer_wv}}\hspace{5mm}
\subfloat[]{\includegraphics[scale = 0.7]{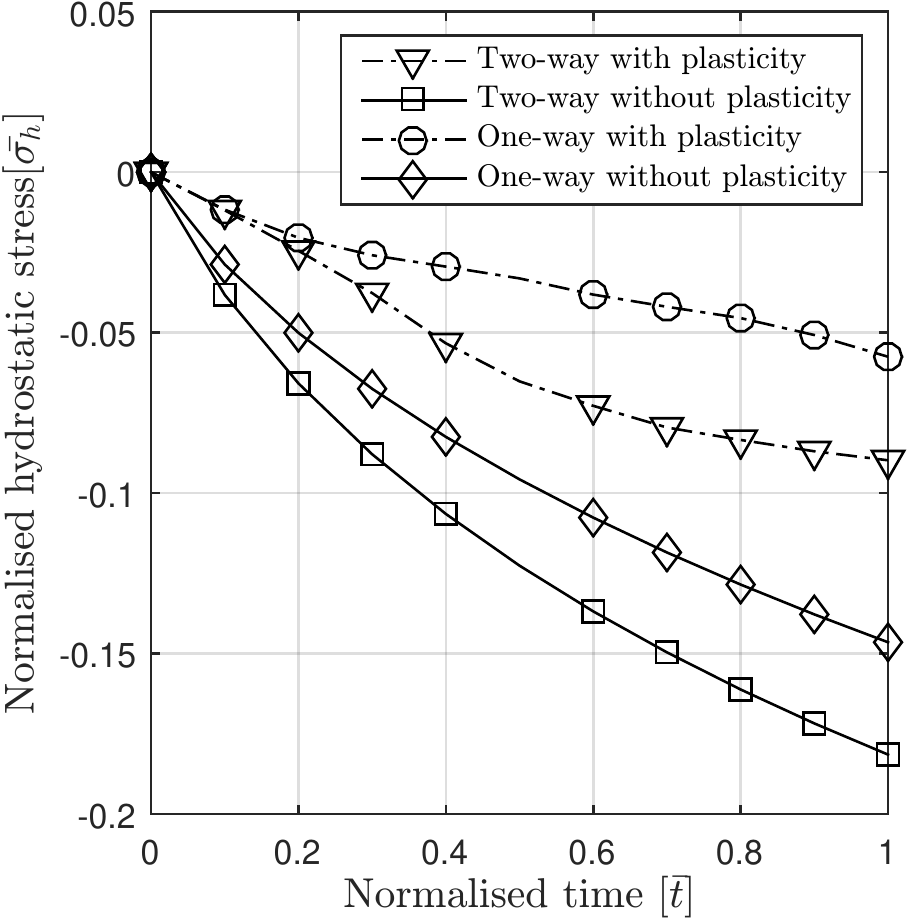}\label{fig:particle_stress_outer_v}}
\caption{ Hydrostatic stress at outer radius (a) without void (b) with void }
\label{fig:particle_stress_outer}
\end{figure}

\begin{figure}[H]
\centering
\subfloat[]{\includegraphics[scale = 0.7]{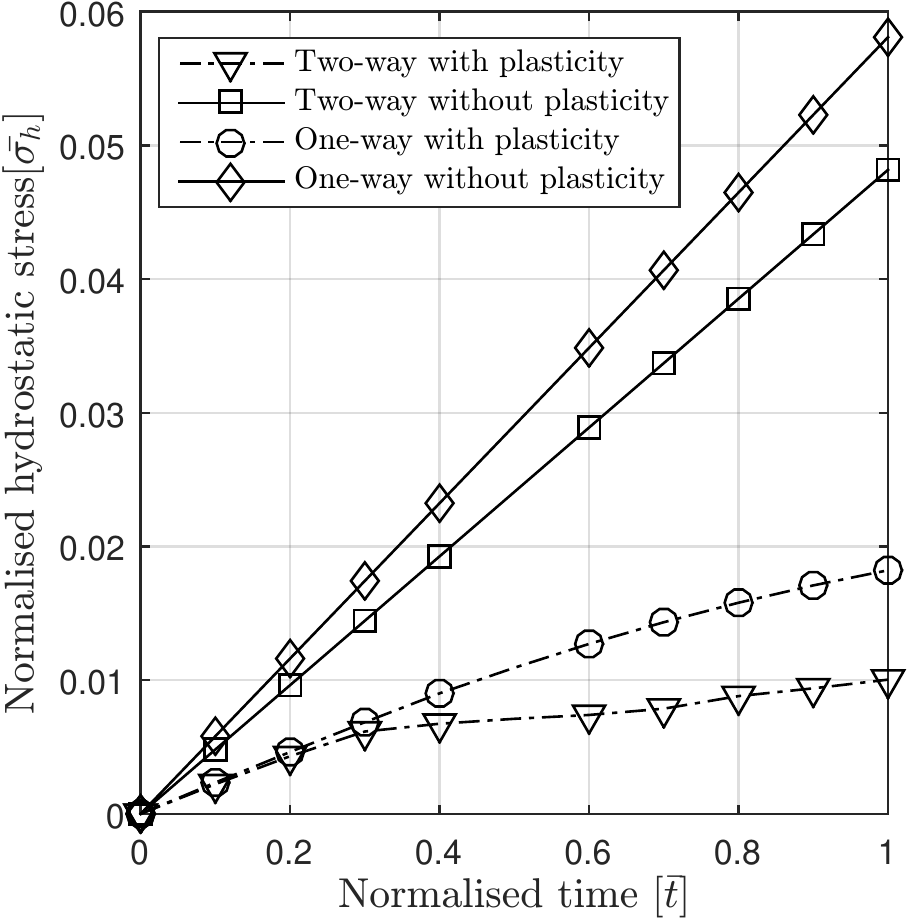}\label{fig:particle_stress_inner_Wv}}\hspace{5mm}
\subfloat[]{\includegraphics[scale = 0.7]{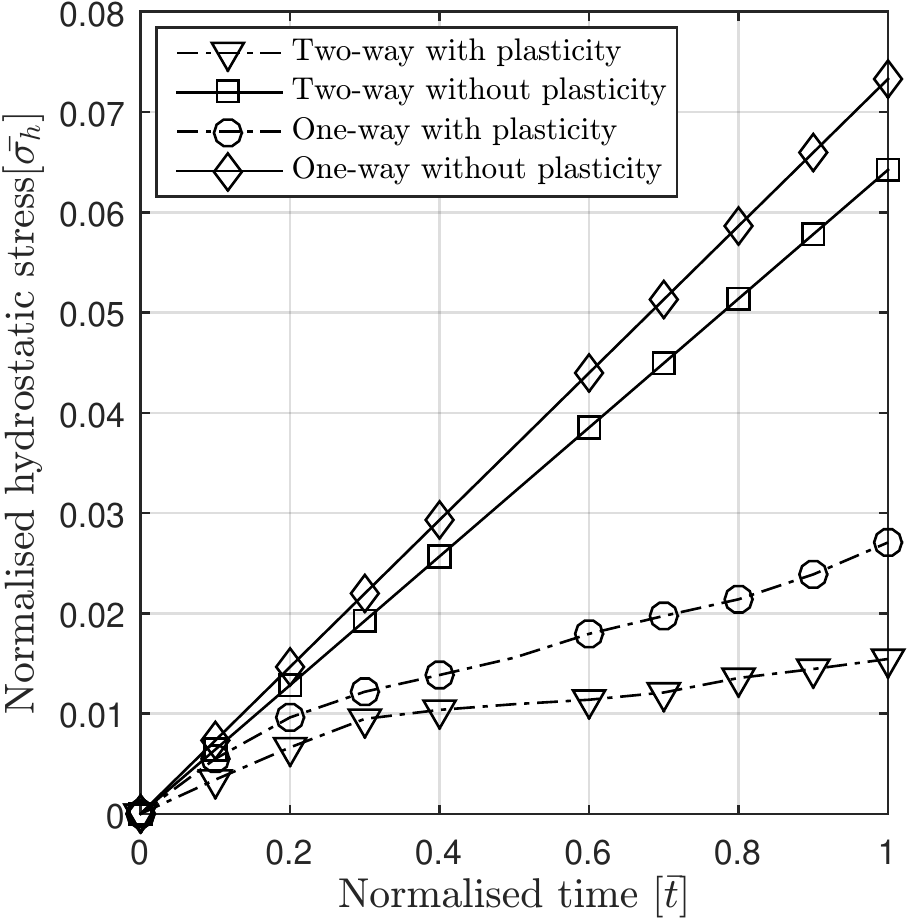}\label{fig:particle_stress_inner_v}}
\caption{ Hydrostatic stress at inner radius (a) without void (b) with void }
\label{fig:particle_stress_inner}
\end{figure}

\begin{figure}[H]
\centering
 \subfloat[]{\includegraphics[scale = 0.7]{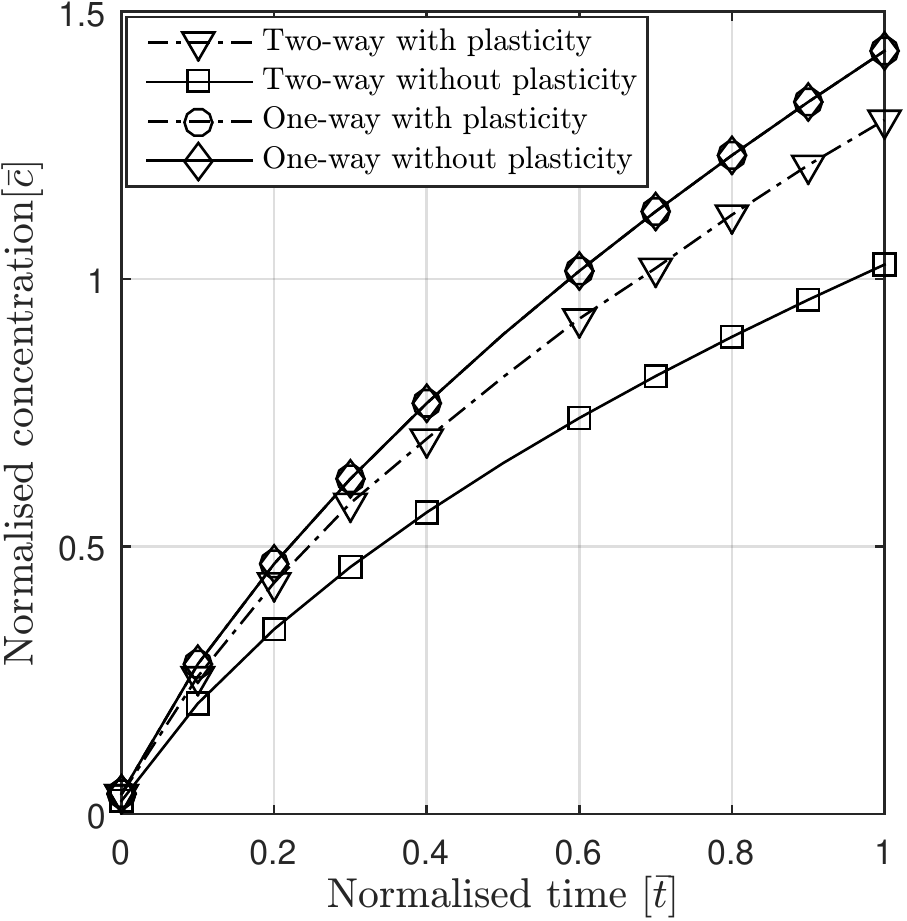}\label{fig:particle_conc_outer_wv}}\hspace{5mm}
\subfloat[]{\includegraphics[scale = 0.7]{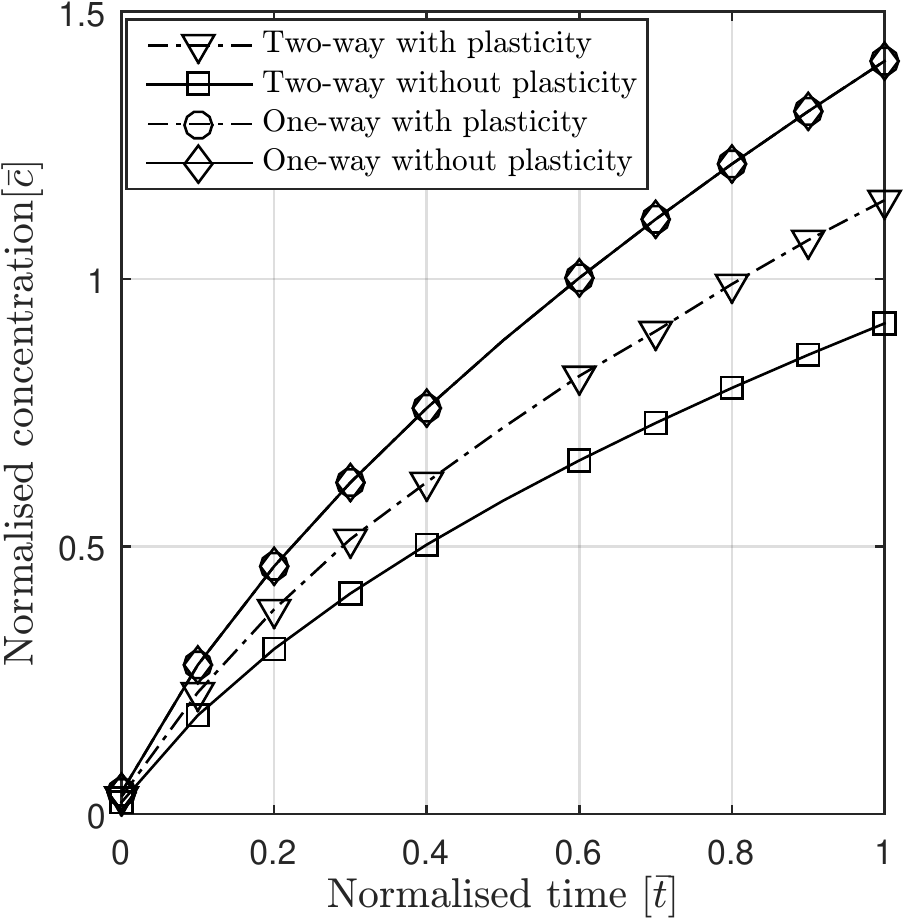}\label{fig:particle_conc_outer_v}}
\caption{ Concentration at outer radius (a) without void (b) with void }
\label{fig:particle_conc_outer}
\end{figure}

\begin{figure}[H]
\centering
 \subfloat[]{\includegraphics[scale = 0.7]{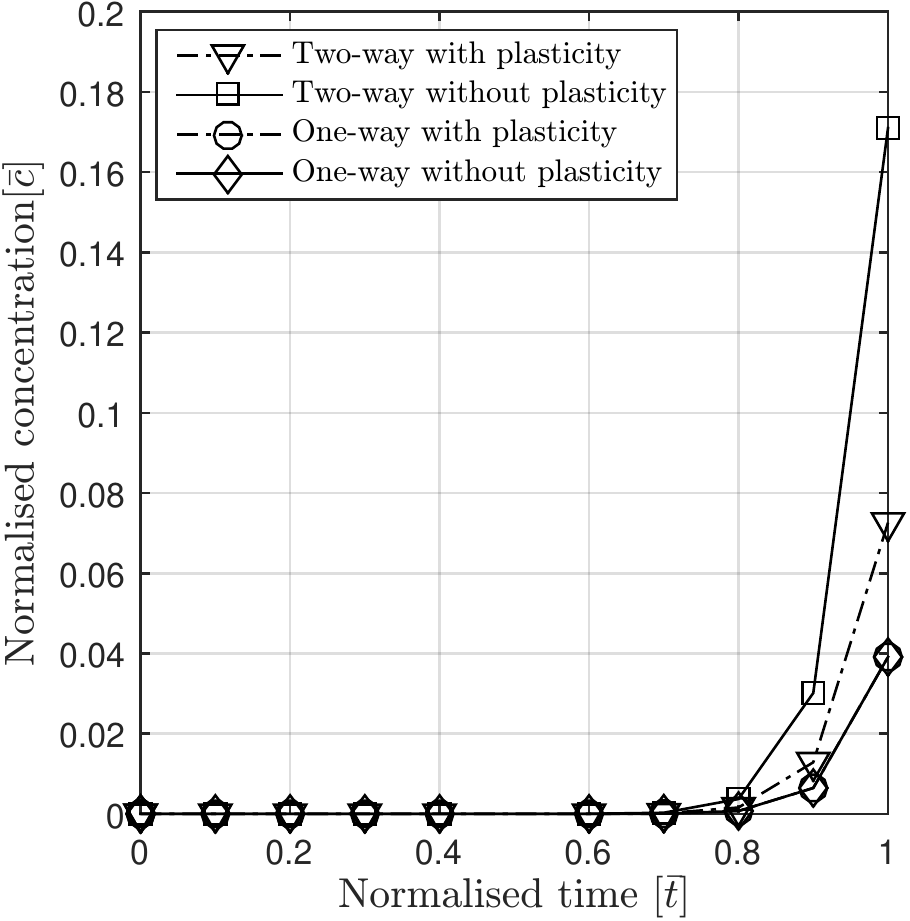}\label{fig:particle_conc_inner_wv}}\hspace{5mm}
\subfloat[]{\includegraphics[scale = 0.7]{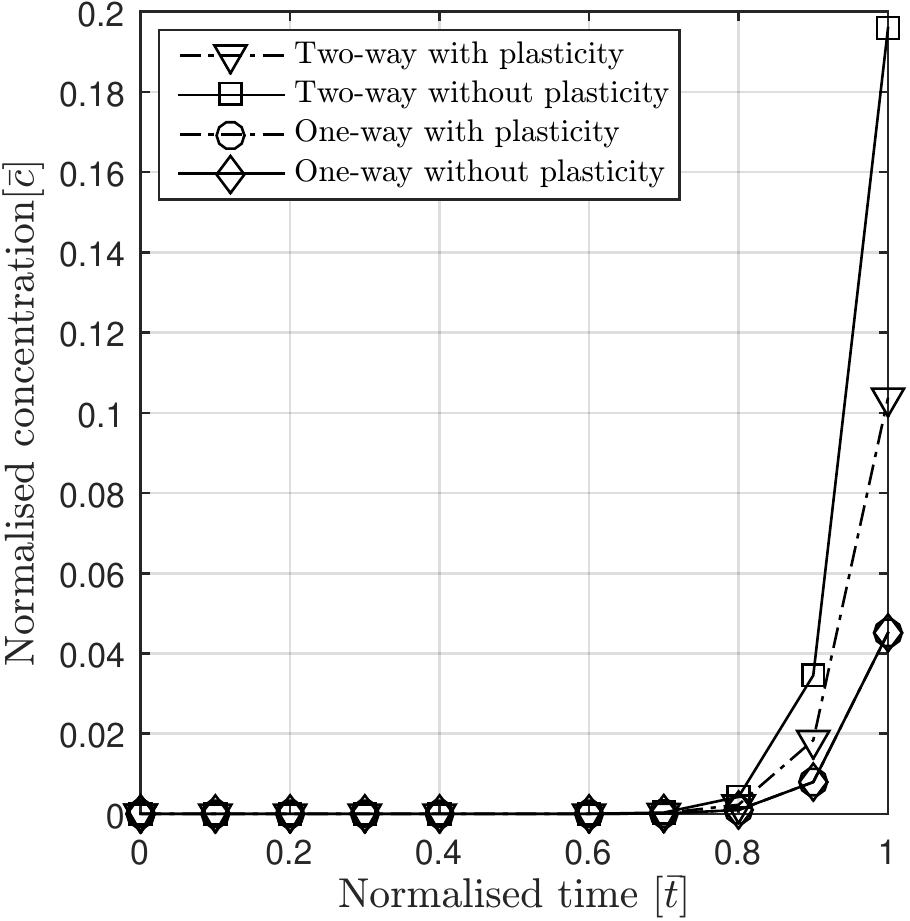}\label{fig:particle_conc_inner_v}}
\caption{ Concentration at inner radius (a) without void (b) with void }
\label{fig:particle_conc_inner}
\end{figure}
Due to the incoming flux (charging operation) on the boundary, the outer surface of the particle undergoes compression and the inner surface of the particle undergo tension, see Figures \ref{fig:particle_stress_outer} and \ref{fig:particle_stress_inner}. From our earlier observation these stresses consequently affects the diffusion process for the two-way coupled system. At the outer radius(compressive region) the two-way coupled system shows lesser concentration of diffusive species as compared to the one-way coupled system whereas at the inner radius (tensile region) the two-way coupled system shows higher concentration as compared to the one-way coupled system, as shown in \frefs{fig:particle_conc_outer} and \ref{fig:particle_conc_inner}. 

\begin{figure}[H]
\centering
\subfloat[]{\includegraphics[scale = 0.7]{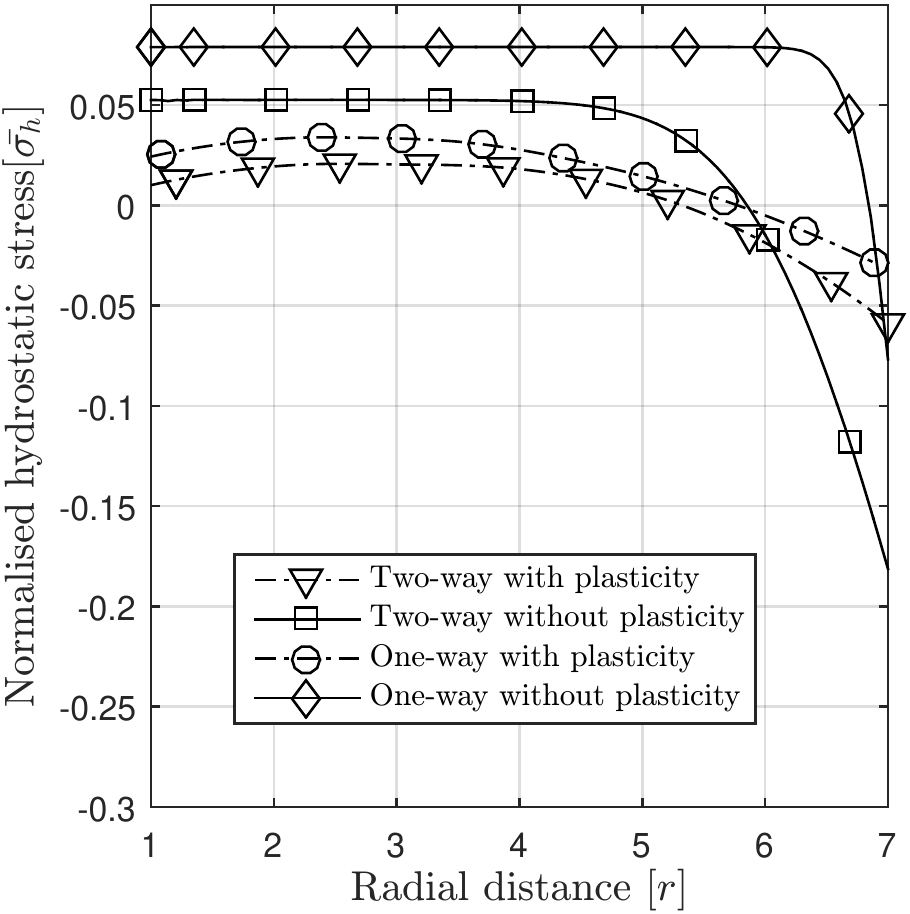}
\label{fig:particle_stress_alongr_wv}
}\hspace{5mm}
\subfloat[]{\includegraphics[scale = 0.7]{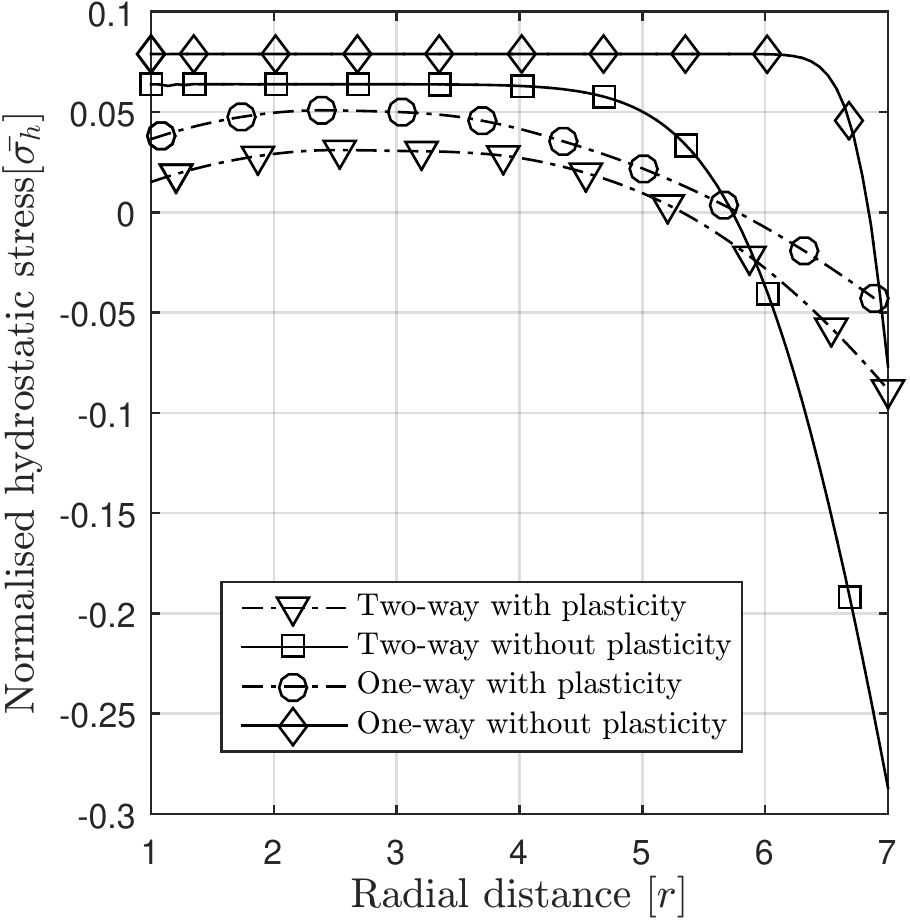}
\label{fig:particle_stress_alongr_v}}
\caption{ Hydrostatic stress along radius (a) without void (b) with void }
\label{fig:particle_stress_alongr}
\end{figure}

The stress in elastoplastic medium is lowered than in pure elastic medium due to plastic yielding. The plastic yielding lowers the tensile stress at the inner radius (tensile region) and hence causes an outflux leading to a decrease in concentration compared to pure elastic media, see \fref{fig:particle_conc_inner}. The inverse effect occurs at the outer radius (compressive region), where a reduction in compressive stress due to plastic yielding causes an influx of diffusing species leading to an increase in concentration, see \fref{fig:particle_conc_outer}. 

\begin{figure}[H]
\centering
\subfloat[]{\includegraphics[scale = 0.7]{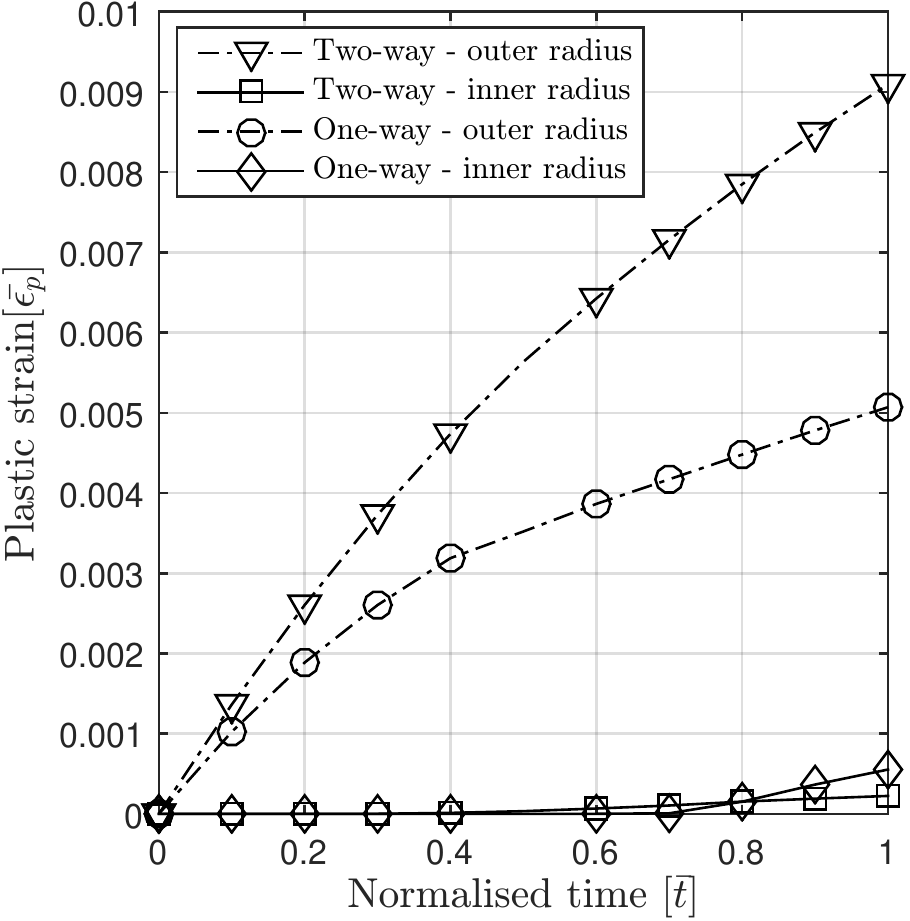}\label{fig:particle_ps_wv}
}\hspace{5mm}
\subfloat[]{\includegraphics[scale = 0.7]{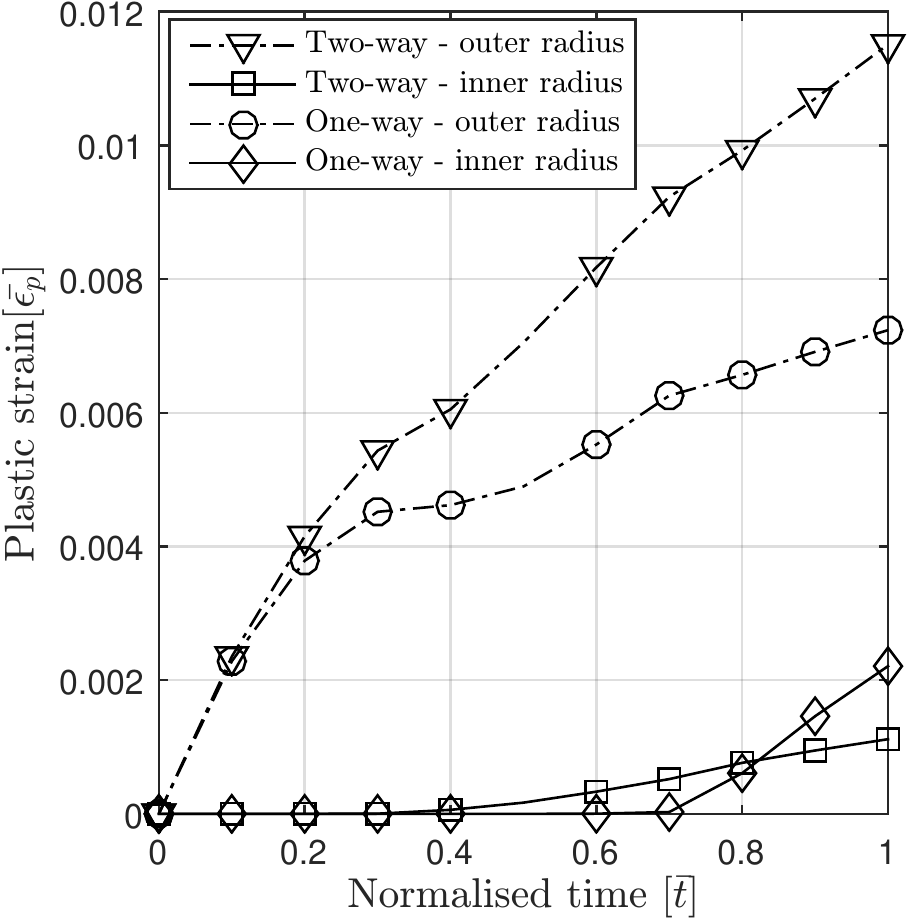}\label{fig:particle_ps_v}}
\caption{ Plastic strain at outer and inner radius (a) without void (b) with void }
\label{fig:particle_ps}
\end{figure}

\begin{figure}[!htbp]
\centering
\subfloat[ ]{
  \includegraphics[scale = 0.5]{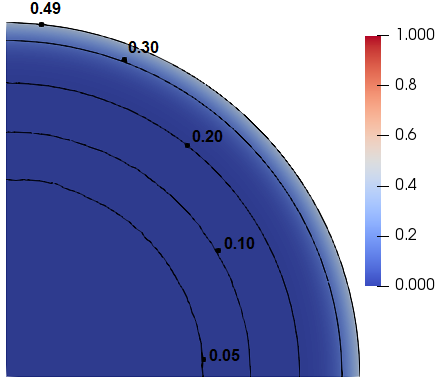}
 } 
 \subfloat[ ]{
 \includegraphics[scale = 0.5]{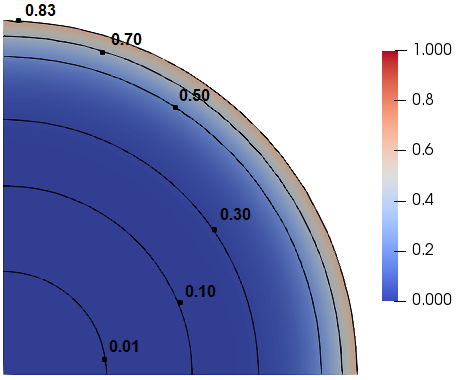}
 } 
\subfloat[ ]{
 \includegraphics[scale = 0.5]{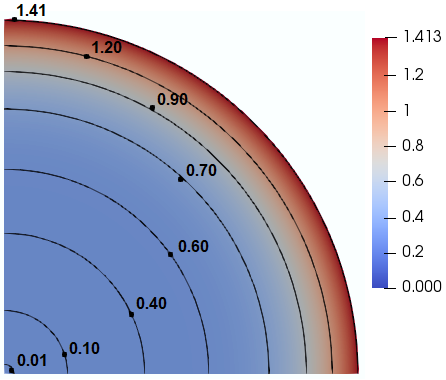}
 } 
\caption{One way coupling - Concentration contours at normalized time (a)  0.2s (b) 0.5s (c) 1s}
\label{fig:particle_oneway_conc_contour_wv}

\end{figure}

\begin{figure}[!htbp]
\centering
\subfloat[ ]{
  \includegraphics[scale = 0.5]{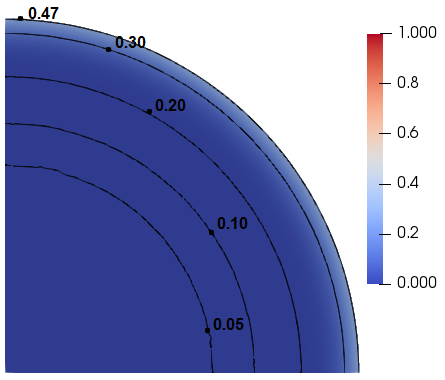}
 } 
 \subfloat[ ]{
 \includegraphics[scale = 0.5]{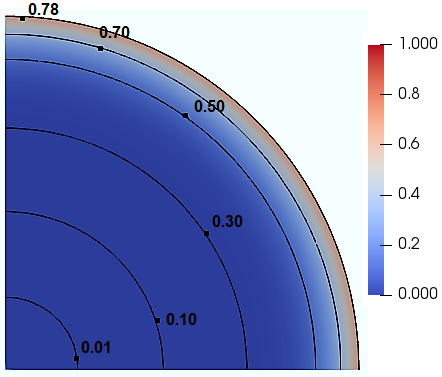}
 } 
\subfloat[ ]{
 \includegraphics[scale = 0.5]{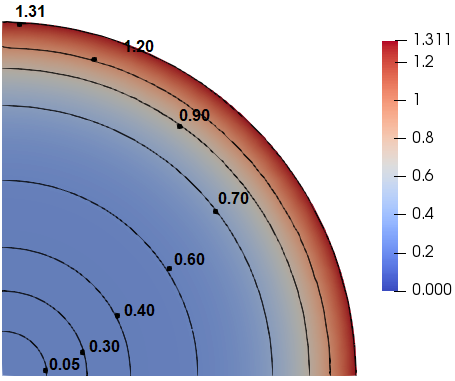}
 } 
\caption{Two way coupling - Concentration contours at normalized time (a)  0.2s (b) 0.5s (c) 1s}
\label{fig:particle_twoway_conc_contour_wv}

\end{figure}

\begin{figure}[H]
\centering
\subfloat[ ]{
  \includegraphics[scale = 0.5]{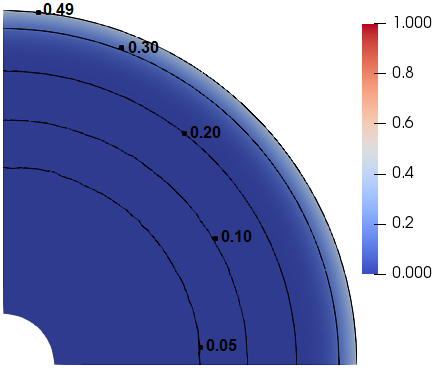}
 } 
 \subfloat[ ]{
 \includegraphics[scale = 0.5]{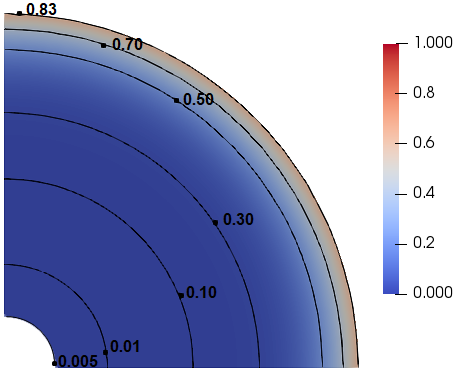}
 } 
\subfloat[ ]{
 \includegraphics[scale = 0.5]{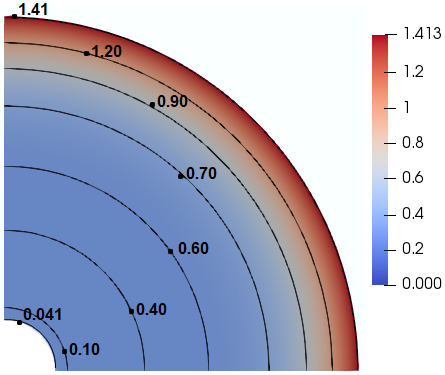}
 } 
\caption{One way coupling - Concentration contours at normalized time (a)  0.2s (b) 0.5s (c) 1s}
\label{fig:particle_oneway_conc_contour_v}

\end{figure}

\begin{figure}[H]
\centering
\subfloat[ ]{
  \includegraphics[scale = 0.5]{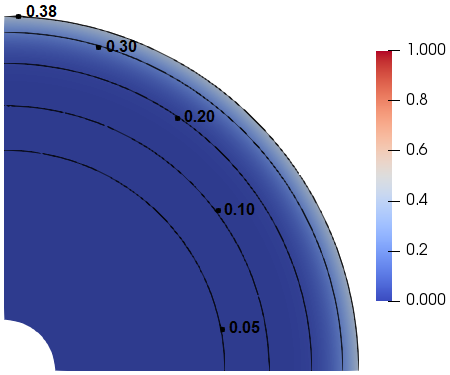}
 } 
 \subfloat[ ]{
 \includegraphics[scale = 0.5]{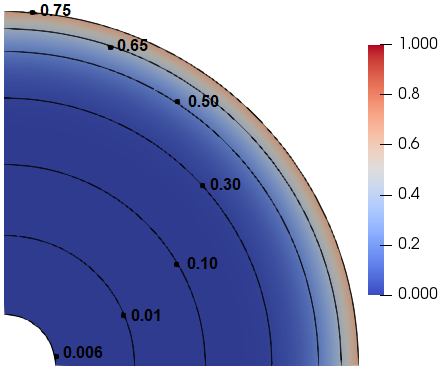}
 } 
\subfloat[ ]{
 \includegraphics[scale = 0.5]{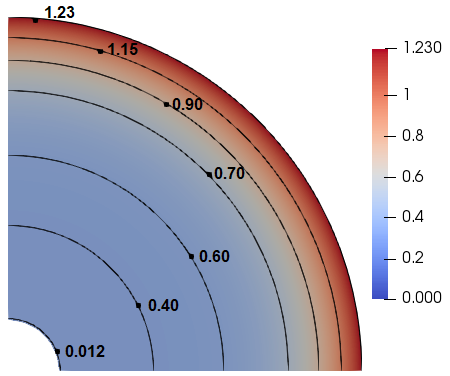}
 } 
\caption{Two way coupling - Concentration contours at normalized time (a)  0.2s (b) 0.5s (c) 1s}
\label{fig:particle_twoway_conc_contour_v}

\end{figure}

When comparing the concentration in one-way coupled model with and without plasticity as shown in  \frefs{fig:particle_conc_outer} and \ref{fig:particle_conc_inner}, we see negligible difference because the stress field does not influence the diffusion process in that case. 
We also note that the compressive stress (at the outer radius) in case of two-way coupled system is more than the one-way coupled system and the tensile stress (at the inner radius) is less in case of two-way coupled system is more than the one-way coupled system, as shown in \frefs{fig:particle_stress_outer} and \ref{fig:particle_stress_inner}. This is due to the stress-relaxation effect caused by flux of diffusing species out of the compressive region and into the tensile region respectively as discussed in the previous section. The plastic yielding reduces the overall singularity in stress due to compression and tension in both one-way and two-way coupled systems. However due to the stress-relaxation effect in two-way coupled system we see (in figure \ref{fig:particle_ps}) higher plastic yielding at the outer radius and lower plastic yielding at the inner radius for two-way coupled system when compared to one-way coupled system. 
From the concentration contour plots as shown in \frefs{fig:particle_oneway_conc_contour_wv}, \ref{fig:particle_twoway_conc_contour_wv}
\ref{fig:particle_oneway_conc_contour_v},
\ref{fig:particle_twoway_conc_contour_v}, concentration is lesser at the outer radius and more at the inner radius in case of the two-way coupled system when compared to one-way system which further illustrates the stress-relaxation effect in two-way system. Along with this, \fref{fig:particle_stress_alongr} gives us a qualitative understanding about the effect of induced stress on diffusion. We observe that due to a high gradient in concentration at the outer radius, the hydrostatic stress and its gradient is also high. As we move closer to the centre the concentration gradient reduces, leading to a smaller/no gradient in the hydrostatic stress. This is similar to effect observed by Swaminathan {\it et. al.,} \cite{Swaminathan2016}.

In order to illustrate the influence of discontinuities, we introduce a void in the particle model. We observe that introduction of the void in the coupled diffusion-deformation process increases the compressive stress at the outer radius (see, \fref{fig:particle_stress_outer}) and the tensile stress at the inner radius, see \fref{fig:particle_stress_inner}. This consequently reflects on the plastic yielding and affects the diffusion process and vice versa, see \fref{fig:particle_ps}. At the outer radius (compressive region) the two-way coupled system shows a reduction in the concentration due to an increase in compressive stress (see, \fref{fig:particle_conc_outer}). At the inner radius (tensile region) the two-way coupled system shows an increase in concentration due to an increase in compressive stress (see, \fref{fig:particle_conc_inner}). The effect of discontinuity on the concentration is negligible in one-way coupled model as expected because the stress does not influence the diffusion process. The concentration contour plots (see, \frefs{fig:particle_oneway_conc_contour_wv}, \ref{fig:particle_twoway_conc_contour_wv}
\ref{fig:particle_oneway_conc_contour_v},
\ref{fig:particle_twoway_conc_contour_v}) further provide evidence to illustrate this effect.

%%%%%%%%%%%%%%%%%%%%%%%%%%%%%%%%%%%%%%%%%%%%%%%%%%

\section{Concluding remarks}
\label{S:5}
In this paper, the effect of stress diffusion interactions in an elasto-plastic material with/without discontinuity using a coupled chemo-mechanics system has been investigated. Moreover, the effect of the fully coupled system over the one-way coupled system is studied systematically for the general framework of the species diffusion in the solid (hydrogen diffusion in steel as well as Li ion diffusion the LIBs). The concentration buildup in the one-way coupled system only depends on the distance from the source where as in two-way coupling the concentration build up depends not only on the distance but the local state of the stress. Study reveals also that the state of stress in the domain is the determining factor in studying the chemo-mechanics system and may change the concentration buildup point (depends on tensile and compressive site). The plastic yielding at the tensile sites reduces the local diffusion species and same affects in reverse for the compressive sites. Moreover, the discontinuities in the domain severely affects the stress-diffusion interactions. 

\section*{Appendix A: non-dimensionalization of equations}
\noindent Consider the balance of diffusion species \erefm{eq1}, 
\begin{equation}
    \frac{\partial c}{\partial t}  - D \nabla_{x} c  + \frac{D \Omega}{\rm{RT}} c \nabla_{x} \boldsymbol{\sigma}_h = 0.
    \label{A2}
\end{equation}

\noindent The \eref{A2} is non-dimensionalized by assuming a new set of dimensionless variables for space ($L^*$), time ($t^*$), concentration ($c^*$) and the hydrostatic stress ($\boldsymbol{\sigma}_h^*$) :
\begin{equation}
 \hat{x} = \frac{x}{L^*}, \,   
\hat{t} = \frac{t}{t^*}, \, 
\hat{c} = \frac{c}{c^*}, \, 
\hat{\boldsymbol{\sigma}}_h = \frac{\boldsymbol{\sigma}_h}{\boldsymbol{\sigma}^*_h}.
\label{A1}
\end{equation}
Now, using the chain rule and substituting new variables in terms of ($\cdot^*$) in \erefm{A2} one can get,
\begin{equation}
    \frac{c^*}{t^*} \frac{\partial \hat{c}}{\partial \hat{t}} -  \frac{c^*}{{L^*}^2} \nabla_{\hat{x}} c + \frac{c^*}{{L^*}^2} \frac{D \Omega}{\rm{RT}} c^* \hat{c} \nabla_{x} \boldsymbol{\sigma}_h = 0.
    \label{A3}
\end{equation}
\noindent Now multiply $t^*$ in \erefm{A3},
\begin{equation}
    c^* \frac{\partial \hat{c}}{\partial \hat{t}} -  \frac{c^* t^* }{{L^*}^2} \nabla_{\hat{x}} c + \frac{c^* t^*}{{L^*}^2} \frac{D \Omega}{\rm{RT}} c^* \hat{c} \nabla_{x} \hat{ \boldsymbol{\sigma}_h} = 0.
    \label{A4}
\end{equation}
On assuming,  
$c^* = 1$, $\frac{D t^*}{{L^*}^2} = 1$, $\dfrac{ \Omega \boldsymbol{\sigma}_h}{RT} = 1$ and upon back substitution in \erefm{A1} we get
$\hat{t} = t \frac{l^2}{D}$, $\hat{c} = \frac{c}{c_{\rm{max}}}$, $\hat{\boldsymbol{\sigma}}_h = \boldsymbol{\sigma}_h \dfrac{\rm{RT}}{\Omega}$. 
\\
Hence, non-dimensionalized equation is given as,
\begin{equation}
    \dfrac{\partial \hat{c}}{\partial \hat{t}}  - \nabla_{\hat{x}} \hat{c}  + \hat{c} \nabla_{\hat{x}} \hat{\boldsymbol{\sigma}}_h = 0. 
\end{equation}

\section*{Appendix B: incremental equations for numerical scheme}
In this appendix we see the numerical procedure for coupled deformation-diffusion using a Kinematic hardening model. Kinematic hardening model uses a back stress, $\beta_{ij}$ , which results in a dependence on loading in the hardening model.

The yield condition in case of Kinematic hardening can rewritten as:
\begin{equation}
    f = \frac{1}{2}(\bm{S}_{ij}-\bm{\beta}_{ij})(\bm{S}_{ij}-\bm{\beta}_{ij}) - \frac{1}{3}\sigma_y^2
\end{equation}

where $\bm{S}_{ij}=\bm{\sigma}_{ij}-\frac{1}{3}\sigma_{kk}\bm{\delta}_{ij}$ is the deviatoric stress. We aim at modelling the back stress, $\beta_{ij}$. We use a linear hardening model governed by:
\begin{equation}
    \bm{\dot{\beta_{ij}}} = h\dot{\bm{\varepsilon^p}}_{ij}
    \label{eqn12}
\end{equation}
where h is the hardening constant. Using the associated flow rule,
\begin{equation}
        \bm{\dot{\beta_{ij}}} = h\dot{\lambda}(\bm{\sigma}_{ij}-\bm{\beta}_{ij})
\end{equation}
Making use of the consistency condition,
\begin{equation}
    \dot{f}=0=\frac{\partial f}{\partial \bm{\sigma}_{ij}}\dot{\bm{\sigma}_{ij}} + \frac{\partial f}{\partial \bm{\beta}_{ij}}\dot{\bm{\beta}_{ij}} = (\bm{S}_{ij}-\bm{\beta}_{ij})\dot{\bm{S}_ij}-(\bm{S}_{ij}-\bm{\beta}_{ij})\dot{\bm{\beta}_{ij}}
\end{equation}
\begin{equation}
     (\bm{S}_{ij}-\bm{\beta}_{ij})\dot{\bm{S}_ij}=(\bm{S}_{ij}-\bm{\beta}_{ij})\dot{\bm{\beta}_{ij}}
\end{equation}
\begin{equation}
    (\bm{S}_{ij}-\bm{\beta}_{ij})\dot{\bm{S}_{ij}}=h\dot{\lambda}(\bm{S}_{ij}-\bm{\beta}_{ij})(\bm{\sigma}_{ij}-\bm{\beta}_{ij})
\end{equation}
\begin{equation}
    \dot{\lambda} = \frac{(\bm{S}_{ij}-\bm{\beta}_{ij})\dot{\bm{S}_{ij}}}{h(\bm{S}_{ij}-\bm{\beta}_{ij})(\bm{\sigma}_{ij}-\bm{\beta}_{ij})}=\frac{(\bm{S}_{ij}-\bm{\beta}_{ij})\dot{\bm{S}_{ij}}}{\frac{2h}{3}\sigma_y^2}
\end{equation}

In order to eliminate the rate of stress, we use again Hooke’s law,
\begin{equation}
    \dot{\lambda}=\frac{(\bm{S}_{ij}-\bm{\beta}_{ij}) \mathbb{C}_{ijkl} (\dot{\bm{\varepsilon}_{kl}} - \dot{\bm{\varepsilon^p}_{kl}}-\dot{\bm{\varepsilon^c}_{kl}})}{\frac{2h}{3}\sigma_y^2}; \; \; \; \;
    \dot{\bm{\varepsilon^p}_{ij}} = \dot{\lambda}(\bm{S}_{ij}-\bm{\beta_{ij}});\; \; \; \;
    \dot{\bm{\beta}_{ij}} = h \dot{\bm{\varepsilon^p}_{ij}}
\end{equation}
Solving first two equations,

\begin{equation}
     \dot{\lambda}=\frac{(\bm{S}_{ij}-\bm{\beta}_{ij}) \mathbb{C}_{ijkl} (\dot{\bm{\varepsilon}_{kl}} - -\dot{\bm{\varepsilon^c}_{kl}})}{\frac{2h}{3}\sigma_y^2 + (\bm{S}_{ij}-\bm{\beta}_{ij}) \mathbb{C}_{ijkl}(\bm{S}_{kl}-\bm{\beta}_{kl})}   \label{eqn13}
\end{equation}

\begin{equation}
    \dot{\bm{\varepsilon^p_{ij}}} = \dot{\lambda}(\bm{S}_{ij}-\bm{\beta_{ij}})
    \label{eqn14}
\end{equation}

\begin{equation}
    \dot{\bm{\sigma}_{ij}}=\mathbb{C}_{ijkl}(\dot{\bm{\varepsilon}_{kl}} - \dot{\bm{\varepsilon^p}_{kl}}-\dot{\bm{\varepsilon^c}_{kl}})
    \label{eqn15}
\end{equation}

In order to implement the constitutive equation, time discretization is used,
\begin{equation}
    \dot{\bm{\sigma}_{ij}}=\frac{\partial \bm{\sigma}_{ij}}{\partial t} = \frac{ \bm{\sigma}_{ij} - \bm{\sigma}^0_{ij}}{\Delta t}
\end{equation}
\begin{equation}
    \bm{\sigma}_{ij} = \bm{\sigma}^0_{ij} + \Delta t \; \dot{\bm{\sigma}_{ij}}; \; \; \;
    \bm{\beta}_{ij} = \bm{\beta}^0_{ij} + \Delta t \; \dot{\bm{\beta}_{ij}}
     \label{eqn16}
\end{equation}

We use the value of stress from the previous time step and approximate the rate of stress using equations \ref{eqn12}, \ref{eqn13}, \ref{eqn14}, \ref{eqn15} to find the current stress states using equation \ref{eqn16} as detailed in procedure~(see, Algorithm \ref{procedure}). For small time increments the numerical solution is accurate and the computational
time is reasonable

%\section*{Reference}
\bibliographystyle{unsrt}
\bibliography{template}

\end{document}